\documentclass[fleqn,usenatbib]{mnras}

\usepackage{newtxtext,newtxmath}

\usepackage[T1]{fontenc}

\DeclareRobustCommand{\VAN}[3]{#2}
\let\VANthebibliography\thebibliography
\def\thebibliography{\DeclareRobustCommand{\VAN}[3]{##3}\VANthebibliography}

\usepackage{graphicx}	
\usepackage{amsmath}	
\usepackage{float}
\usepackage{natbib}
\usepackage{caption}
\usepackage{gensymb}
\usepackage{bm}
\usepackage[output-decimal-marker={,},exponent-product=\cdot]{siunitx}
\usepackage[utf8]{inputenc}
\usepackage{newtxtext,newtxmath}
\usepackage{hyperref}

\newcommand{\mr}{\mathrm}

\title[AGN jet-inflated bubbles]{The complex interplay of AGN jet-inflated bubbles and the intracluster medium}

\author[F. Huško \& C. G. Lacey]{
Filip Huško,$^{1}$\thanks{E-mail: filip.husko@durham.ac.uk}
Cedric G. Lacey$^{1}$
\\
$^{1}$ Institute for Computational Cosmology, Department of Physics, University of Durham, South Road, Durham, DH1 3LE, UK\\
}

\date{Accepted XXX. Received YYY; in original form ZZZ}

\pubyear{2022}

\begin{document}

\label{firstpage}
\pagerange{\pageref{firstpage}--\pageref{lastpage}}
\maketitle

\begin{abstract}
We use SWIFT, a smoothed particle hydrodynamics code, to simulate the evolution of bubbles inflated by active galactic nuclei (AGN) jets, as well as their interactions with the ambient intracluster medium (ICM). These jets inflate lobes that turn into bubbles after the jets are turned off (at $t=50$ Myr). Almost all of the energy injected into the jets is transferred to the ICM very quickly after they are turned off, with roughly $70\%$ of it in thermal form and the rest in kinetic. At late times ($t>500$ Myr) we find the following: 1) the bubbles draw out trailing filaments of low-entropy gas, similar to those recently observed, 2) the action of buoyancy and the uplift of the filaments dominates the energetics of both the bubbles and the ICM and 3) almost all of the originally injected energy is in the form of gravitational potential energy, with the bubbles containing $15\%$ of it, and the rest contained in the ICM. These findings indicate that feedback proceeds mainly through the displacement of gas to larger radii. We find that the uplift of these filaments permanently changes the thermodynamic properties of the ICM by reducing the central density and increasing the central temperature (within $30$ kpc). We propose that jet feedback proceeds not only through the heating of the ICM (which can delay cooling), but also through the uplift-related reduction of the central gas density. The latter also delays cooling, on top of reducing the amount of gas available to cool.
\end{abstract}






\section{Introduction}

The majority of baryonic matter in galaxy clusters is located in the intracluster medium (ICM), a diffuse and hot (X-ray emitting; $T>10^7$ K) gaseous halo that roughly traces the dark matter distribution (\citealt{Pratt2009}, \citealt{Sun2009}, \citealt{Lin2012}). Observations of the ICM in X-rays reveal the existence of cavities, regions marked by a lack of X-ray emission (e.g. \citealt{Birzan2004}, \citealt{McNamara2005}, \citealt{Wise2007}). These cavities are coincident with bubbles inflated by jets of relativistic particles, which are visible in radio frequencies due to synchrotron emission (\citealt{Blandford1979}, \citealt{Urry1995}). The jets are launched from active galactic nuclei (AGN), hosting the accreting supermassive black holes (SMBH) of the central galaxies of the clusters (\citealt{Ghisellini1993}, \citealt{Biretta1999}). These jets are decelerated by the ICM and deposit their energy into it (\citealt{McNamara2007}, \citealt{Fabian2012}, \citealt{McNamara2012}).

The central region of the ICM (inner few hundred kpc) is dense and cool enough that the cooling time is often significantly shorter than the Hubble time (\citealt{Hudson2010}). As a result, we would expect the central galaxy of many galaxy clusters to harbour very high rates of cool gas deposition and star formation. However, other than a few exceptions (\citealt{Odea2008}, \citealt{McDonald2015}), this is usually not the case (\citealt{Edge1991}, \citealt{Fabian1994}, \citealt{McDonald2018}). Furthermore, observations of emission lines due to cool or cooling gas, in X-rays (e.g. \citealt{Peterson2003}) and optical (e.g. \citealt{Edge2002}), infrared (e.g. \citealt{Odea2008}) and ultraviolet wavelengths (e.g. \citealt{Bregman2006}), are consistent with low cooling rates. The central galaxies of galaxy clusters are typically 'red and dead', like most other massive elliptical galaxies (\citealt{Wiklin1995}, \citealt{Salim2007}, \citealt{Young2011}, \citealt{Whitaker2012}, \citealt{Davis2019}). 

In order to keep the central galaxies of galaxy clusters devoid of significant amounts of cool gas and star formation, AGN jets have been proposed as a heating mechanism that counters the radiative cooling of the ICM (\citealt{Fabian2012}, \citealt{Werner2019}, \citealt{Eckert2021}). The power required to create the X-ray cavities (a proxy for the jet power) has been found to be correlated with the X-ray luminosity of the ICM (\citealt{Rafferty2006}, \citealt{Nulsen2009}, \citealt{Hlavacek-Larrondo2012}, \citealt{Russell2013}). It is also sufficient to offset cooling, indicating that AGN feedback in the form of relativistic jets is a plausible mechanism of star formation quenching, by depriving the central galaxies of the required cool gas. 

Early idealised simulations of single-episode AGN jet feedback often circumvented the highly-uncertain jet physics, and instead manually placed bubbles of hot gas into the ICM (\citealt{Churazov2001}, \citealt{Bruggen2002}, \citealt{Reynolds2006}, \citealt{Pavlovski2008}, \citealt{Sternberg2008}). The bubbles were then studied as they interacted with and rose through the ICM on account of buoyancy. More recent simulations of single-episode jet feedback have focused mostly on the active phase of the jets. This has been possible due to significant improvements in the reliability of such simulations, which are a result of our better understanding of jet physics and numerical modeling, as well as improved computational capabilities. Such simulations may include only hydrodynamical aspects of the jets (e.g. \citealt{Komissarov1998}, \citealt{Hardcastle2013}, \citealt{Horton2020}). However, other aspects are often included, such as: relativistic physics (e.g. \citealt{Walg2013}, \citealt{English2016}, \citealt{Choi2017}), magnetic fields (e.g. \citealt{Hardcastle2014}, \citealt{Tchekhovskoy2016}, \citealt{Mukharjee2020}), radiative cooling (e.g. \citealt{Blondin1990}, \citealt{Stone1997}, \citealt{Guo2018}) or cosmic rays (e.g. \citealt{Guo2011}, \citealt{Ehlert2018}, \citealt{Yang2019}). The main focus of such studies is often the jet energetics, i.e. how much energy is transferred to the ICM, where, how quickly and in what form (\citealt{Morsony2010}, \citealt{Bourne2017}, \citealt{Weinberger2017}, \citealt{Bourne2019}), as well as by what means (\citealt{Perucho2010}, \citealt{Bambic}, \citealt{Yang2019}, \citealt{Wang2022}).  

In \cite{Husko2022a} we simulated the interactions of AGN jets with the ambient medium using the SWIFT code (\citealt{Schaller2018}) and its its smoothed particle hydrodynamics (SPH) implementation (\citealt{Borrow2022}). These simulations represent the first hydyodynamical tests of AGN jets performed with the SPH method. We focused on a simple set-up where constant-power jets were launched into a constant-density ambient medium, with the aim of reaching the self-similar regime of evolution (e.g. \citealt{Kaiser2007}). This was done in order to understand the basic features of the jets, the lobes that they inflate and their interactions with the ambient medium. In addition, using such a simple set-up allowed us reliably compare the results of the simulations with theoretical models of jets that inflate self-similar lobes. We found good agreement between our simulations and theoretical predictions.

In this paper, we will perform high-resolution and long-duration (Gyr-scale) simulations of AGN jets and bubbles, all the way from the jet launching phase through bubble inflation and subsequent buoyant evolution. Our simulations are of high-power, explosive feedback, rather than gentle, 'maintenence-mode' feedback. Drag and mixing with the ICM (entrainment) are both likely to be important in the evolution of jet-inflated bubbles (\citealt{Pope}). Observations with ALMA have also found that X-ray cavities/radio bubbles are often enveloped by cool ICM gas (e.g. \citealt{Russell2017}), or accompanied by cool gas filaments trailing them (e.g. \citealt{Russell2016}, \citealt{Vantyghem2018}, \citealt{Olivares2019}, \citealt{Russell2019}). Observations at other wavelengths also find such filaments (e.g. \citealt{Wilman2009}, \citealt{Salome2011}, \citealt{Tremblay2015}, \citealt{Gendron-Marsolais2017}, \citealt{Maccagni2021}, \citealt{Gatuzz2021}). Jet activity is associated with metal outflows (\citealt{Sanders2005}, \citealt{Kirkpatrick2009}, \citealt{Doria2012}); it is likely that the uplift associated with these filaments is responsible for the metal redistribution. These filaments have been successfully reproduced in simulations (\citealt{Revaz2008}, \citealt{Li2014}, \citealt{Brighenti2015}, \citealt{Qiu2019}), but their role in feedback has not been well studied. In \cite{Husko2022b} we studied AGN feedback in simulations of idealised galaxy groups and clusters, where the jets were launched from black holes that grew in a self-consistent manner (based on \cite{Bondi} accretion) and whose spin was realistically evolved. We found that jet-inflated bubbles were ubiquitously followed by the uplift of cool, low-entropy ICM gas that formed filaments.

According to the analytical model proposed by \cite{Pope}, these filaments trailing the bubbles, found in both observations and simulations, form on account of two different processes. The first of these is the Darwin drift (\citealt{Darwin1953}), which constitutes the main body of the filaments. These filaments form and rise on account of displacement by the bubbles: the bubbles push aside some of the ICM, which is then pulled back into the region left empty by the moving bubbles. According to theoretical calculations, the volume associated with the drift should be a constant fraction of the volume of the bubbles, and dependant on the bubble shape (\citealt{Darwin1953}, \citealt{Dabiri2006}). The second process that may play a role in the observed filaments is the wake, which occurs only in the presence of gravity. The wake is associated with the trapping of some of the ambient medium in an indentation at the bottom of a bubble, as the bubble begins to rise due to buoyancy. The wake has been observed in fluidized beds (along with the drift, see \citealt{Yang2003}, \citealt{Crowe2005}), where some of the solids that make up the bed travel upwards at the back of air bubbles. This mass acts as additional inertial mass of the bubble, and it moves at the same velocity, but it does not mix with the bubble. The mass of the wake should remain constant with time once the bubbles begin to rise, and it should be some fraction of the initially displaced mass (\citealt{Pope}).

One of the goals of this paper is to study these secondary processes that occur in tandem with bubble evolution. Our main focus is on the drift and wake, but drag and entrainment also play important roles. To date, no simulation of jet-inflated bubbles in a realistic ICM set-up have been performed with the aim of measuring the masses and volumes of the drift and wake, and comparing those with theoretical expectations and experiments. We expect the drift and wake to be energetically significant (due to their masses and volumes being comparable to that of the bubbles), at least at late times in the simulation, once the bubbles have moved significantly.

The outline of this paper is as follows. In Section \ref{sec:sec1.1}, we discuss the numerical and physical details of our simulations. In Section \ref{sec:gen_char} we discuss some general properties of our simulated jets and bubbles, including their morphology, energetics and impact on profiles of gas properties. In Section \ref{sec:drift_wake} we analyze in detail the properties and effects of important additional physics such as drag and entrainment, but we focus especially on the drift and wake that form behind the rising bubbles. In Section \ref{sec:param_study} we show results on bubbles simulated with varying parameters, including physical, jet-related ones and numerical ones. In Section \ref{sec:conclusions} we summarise and conclude.

\section{Simulations}
\label{sec:sec1.1}

In this section we discuss the numerical code and hydrodynamical scheme that we use to simulate the jets and bubbles. We also discuss the details of the physical set-up that we use in these simulations, as well as the different simulations we perform, with varying parameters.

\subsection{Numerical code and hydrodynamical scheme}

We use the open-access\footnote{\href{https://swift.dur.ac.uk/}{https://swift.dur.ac.uk/}} SWIFT code (\citealt{Schaller2018}), which includes hydrodynamics, gravity, cosmology and many subgrid physical processes such as radiative cooling, star formation, chemistry and feedback from stars and black holes. The default hydrodynamical scheme implemented in SWIFT is SPHENIX (\citealt{Borrow2022}), a smoothed particle hydrodynamics method (SPH; \citealt{Monaghan}). 

Traditional SPH codes are known to suffer from artificial surface tension problems (\citealt{Argetz2007}, \citealt{Sijacki2012}, \citealt{Nelson2013}). SPHENIX includes artificial viscosity, which is necessary in order to capture shocks. SPHENIX also includes artificial conductivity, which helps reduce unwanted surface tension otherwise present in SPH simulations, allowing for mixing between flows that are in pressure equilibrium but contrasting in temperature and/or density. An artificial viscosity limiter is included to prevent spurious viscosity in shear flows, while an artificial conductivity limiter is included to prevent spurious energy transfer in all flows.

\subsection{Physical set-up}

We launch our jets into a spherically symmetric gas distribution, which represents the intracluster medium of a dark matter halo with a virial mass of $M_\mr{200}=10^{14}$ $\mr{M}_\odot$, corresponding to a virial radius of $R_\mr{200}\approx950$ kpc at $z=0$\footnote{The virial radius, $R_{200}$, is calculated such that the mean density of the halo is 200 times the critical density at $z=0$.}. We assume a concentration parameter of $c=5.6$. We do not include dark matter explicitly, and instead model its effects through a fixed external \cite{NFW} (NFW) potential. The gas distribution is modeled using the $\beta$ profile, which has the following form:
\begin{equation}
    \rho(r)=\frac{\rho_0}{[1+(r/r_\mr{c})^2]^{3\beta/2}}
\label{eq:eq1}
\end{equation}
Here, $\rho_0$ is a normalisation constant, $r_\mr{c}$ the core radius, and $\beta$ the parameter that determines the slope of the density profile at large radii. We use $\beta=0.5$, yielding a slope of $-1.5$ at large radii, which is appropriate for a $M_{200}=10^{14}$ $\mr{M}_\odot$ halo (\citealt{Voit2002}). The core radius is set to $25$ kpc, which is also appropriate for such haloes. $\rho_0$ is calculated from the condition that the total gas mass within $R_\mr{200}$ is $8$ per cent of the total dark matter mass (\citealt{Pratt2009}, \citealt{Sun2009}, \citealt{Lin2012}). The actual gas profile extends out to $4R_\mr{200}$. In order to reduce the time required to computationally evolve the system, we increase the gas particle masses progressively as $m_\mr{gas}\propto r^2$ beyond the virial radius. The gas is initially assumed to be in hydrostatic equilibrium, with the external NFW potential being used to calculate the pressure profile from this assumption. We then calculate the temperature profile from the equation of state, assuming the gas is ideal.

As a default, we do not include additional physics such as self-gravity and radiative cooling. We have attempted simulations with both included, as well as realistic rotation in the gaseous halo, but we find that none of these have a significant impact on our simulations, at least for 2 Gyr (the simulation time). 

\subsection{Jet launching}

We launch jets from an initially conical set-up, similar to that in \cite{Husko2022a}. We place two particle reservoirs of conical shapes in opposite directions along the $z-$axis in the centre of the gaseous halo. Each cone is 5 kpc long and is defined by its half-opening angle $\theta_\mr{j}$, the same as the launching angle of the jet. The reservoir particles are launched from the cones with a velocity of $v_\mr{j}$, with the velocity vector pointing radially from the centre. The temperature of the reservoir is the same as the rest of the gas in the centre of the halo ($\approx10^7$ K). The jets are active for $T_\mr{j}=50$ Myr with a (total, summed over both jets) power of $P_\mr{j}$. The particles are kicked at intervals of $(1/2)m_\mr{gas}v_\mr{j}^2/P_\mr{j}$, from the outside in.

\subsection{Seeding of particle positions}
\label{sec:seed_pos}

In our simulations we find that jet-inflated bubbles show instabilities at late times, which grow from random perturbations. As a result, the initial positions of particles launched into the jets from the conical reservoirs can have an impact on the morphology of the bubbles. We have attempted various choices, including: i) random placement within the cones, ii) a uniform, face-centred cubic lattice of particles, out of which cones are cut out, and iii) a hydrodynamical or gravitational glass of particles, out of which, again, we cut out the cones. We find that instabilities and asymmetries arise in all three cases, even with a perfectly symmetric uniform initial set-up. In the simulations presented in this paper, the default choice was to use a hydrodynamical glass.

We have attempted the same three choices when seeding the particles in the gaseous halo, which represents the bulk of the particles being evolved. An additional complication is that with the gaseous halo, we construct the desired spherically symmetric density distribution by radially rescaling the positions of an initially homogenous (constant-density) cube. We find that drawing particle positions randomly results in overdensities and underdensities that can take more than a few Gyr to homogenize. Similarly, using a uniform cubic face-centred lattice results in radial spokes of overdensities and underdensities, which arise from the uniform set-up being radially rescaled and deformed. Our choice is, again, to use a hydrodynamical glass. Even with this choice, there are some perturbations that arise in the gaseous halo, but they are much smaller than in the other two cases.

\subsection{Parameter choices and variations}

Our standard choice for the physical, jet-related parameters is: jet duration $T_\mr{j}=50$ Myr, jet power $P_\mr{j}=3.16\times10^{45}$ ergs$^{-1}$, launching velocity $v_\mr{j}=2\times10^4$ kms$^{-1}$ and half-opening angle $\theta_\mr{j}=15\degree$. These choices are a result of much trial and error, and we make them as they result in the self-similar regime of jet evolution at early times (that features lobes that smoothly transition into bubbles, as opposed to ballistic jets). In addition, the bubbles form at distances of several hundred kpc, allowing the study of the interplay between the bubbles and the ICM as they rise out to the virial radius and beyond. We note that the choice of jet power does not correspond to gentle feedback that would keep the core of the ICM from cooling. Instead, the high power corresponds to what is likely a Fanaroff-Riley type II (FRII) jet episode (\citealt{Fanaroff}), launched from a state of high accretion rate of the SMBH hosted by the central galaxy. Thus, our fiducial jet episode would probably be triggered by an event such as a galaxy merger or disc instability. The subrelativistic velocity we have chosen allows a fairly good resolution to be achieved in the bubbles (see discussion below). However, we do not simulate our jets with transrelativistic velocities that are a significant fraction of $c$, which would correspond to the observed FRII jets (e.g. \citealt{Mullin2009}). While this is technically possible with our simulations, such jets and the lobes they inflate would be resolved with only thousands of particles. We vary all of the parameters mentioned above, with the exception of the jet duration, in order to study their impact on the properties of the bubbles, as well as on their interaction with the ICM.

The main simulation we will discuss in this paper was performed with a numerical resolution of $m_\mr{gas}=10^4$ $\mr{M}_\odot$. The total number of particles in this simulation is $\approx1.4\times10^9$. In the centre of the halo, the typical smoothing length (and thus spatial resolution) is 0.2 kpc. The number of particles injected into each jet in this simulation (given the standard, jet-related parameters discussed above) is $1.35\times10^5$, which is more than sufficient to achieve convergent properties for self-similar jet-inflated lobes (\citealt{Husko2022a}). In this simulation, the jets and subsequent bubbles entrain significant amounts of material (as we will discuss in the results), so the number of particles in each bubble at late times is $\approx10^6$. If one were to compare this resolution to a grid-based code, very crudely this corresponds to a grid of $30\times30\times30$ cells across the bubble, or a spatial resolution of $1-10$ kpc, depending on the evolutionary phase (i.e. size and shape) of the bubbles. We have also performed simulations with resolutions of up to $m_\mr{gas}=10^7$ $\mr{M}_\odot$, in order to test the convergence properties of our simulations. Finally, we also performed simulations where we vary the hydrodynamical scheme used to evolve the bubbles. We compare SPHENIX with a 'minimal SPH' scheme (\citealt{Monaghan}), without any artifical viscosity or conductivity, and with anarchy-pu, the scheme used in the EAGLE simulations (\citealt{Schaye2015}, \citealt{Schaller2015}). We also refer the reader to \cite{Braspenning2022} for a more comprehensive comparison of different numerical codes, SPH and grid-based, in the context of the 'blob test' (the interaction of a dense bubble with a supersonic wind).

\section{General characteristics of jet-inflated bubbles}
\label{sec:gen_char}

In this section we will discuss general features of bubbles that appear in all of our simulations. Some of the details differ from one simulation to another, but in this case we choose to analyze one simulation in detail: our highest resolution simulation (with $m_\mr{gas}=10^4$ $\mr{M}_\odot$). 

\subsection{Jet launching and lobe inflation}

\begin{figure*}
\includegraphics[width=1.01\textwidth, trim = 0 10 0 0]{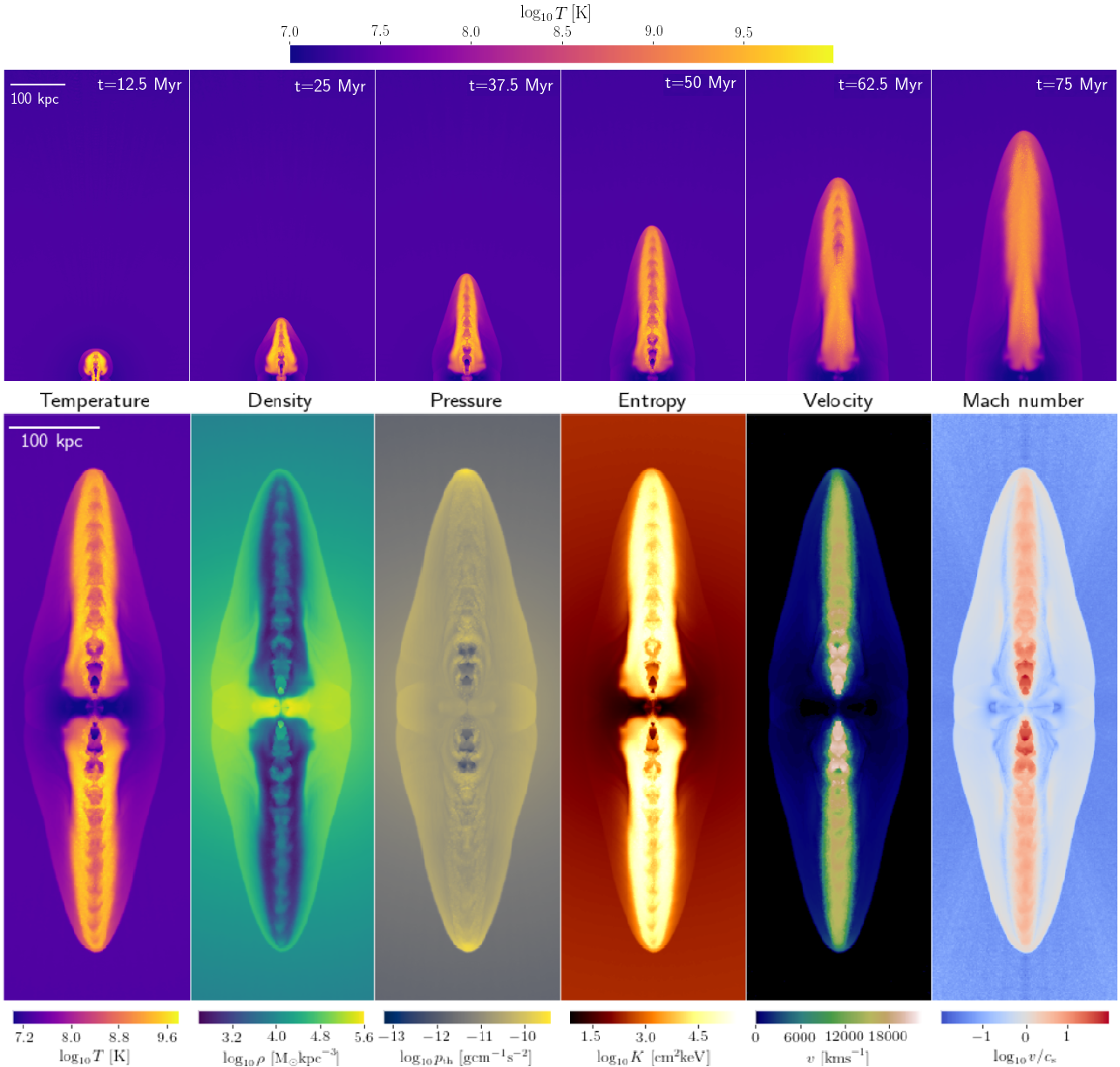}
\caption{Visualisations of jet launching in our highest-resolution simulation ($m_\mr{gas}=10^4$ $\mr{M}_\odot$), through temperature maps at different times (top) and maps of various properties, as labeled (bottom), at $t=50$ Myr, which is when they are turned off. The panels measure $400\times120$ kpc$^2$ and show slices with a depth of $15$ kpc. The jet power is $3.16\times10^{45}$ ergs$^{-1}$. Jet particles are kicked conically within half-opening angles of $15\degree$, and with launching velocities of $2\times10^4$ $\mr{km}\mr{s}^{-1}$. The total gas mass, its distribution and the external potential correspond to a dark matter halo with a virial mass of $M_{200}=10^{14}$ $\mr{M}_\odot$.}
\label{fig:fig1}
\end{figure*}%

The origins of AGN bubbles can be traced to the jets that inflate them, and more directly to the lobes created by the shocking of the gas launched into the jets. These lobes are the precursors of the bubbles, which exist in the phase while the jets are still active. In Fig. \ref{fig:fig1}, in the top panels we show slices of the temperature distribution of gas in the central regions of the gaseous halo, during this initial phase (we show the slices up to $75$ Myr, but the jets are active for $50$ Myr). In the bottom panels we show various properties of the gas in slices at $t=50$ Myr. 

Three distinct features are visible in these plots. The jet spine (or more simply, the jet itself) is made up of the relatively cold and dense outflowing gas that has been kicked as part of the jet launching process, but has not yet been shocked (this is visible through its high velocity and Mach number). Surrounding the jet itself are lobes of very hot ($T\approx10^9$ K) and high-entropy gas (we define the entropy as $K=k_\mathrm{B}T/n^{2/3}$, with $n$ the number density of particles), which is made up of previously shocked jet gas. Finally, the launching of the jet also results in a bow shock propagating through the ICM, which transitions from supersonic velocities near the jet head to expansion at the sound speed (i.e. a sound wave) far away from the jet head.

Our jets and jet-inflated lobes show features that match the theory of self-similar lobes (e.g. \citealt{Kaiser1997} and \citealt{Komissarov1998}). The self-similar regime begins roughly when the swept-up mass of the ambient medium exceeds the mass of the material launched directly into the jets. The jet transitions from being ballistic to experiencing significant shocking, since it has to impart significant amounts of its own momentum to the ambient medium in order to sweep it up. Given a jet power and ICM density, the main parameter that controls where a jet transitions from the ballistic to the self-similar regime is the jet launching velocity, $v_\mr{j}$ (\citealt{Kaiser2007}). Higher values lead to the transition occurring at smaller distances. Our choice ($v_\mr{j}=2\times10^4$ kms$^{-1}$) leads to a transition at $\approx2$ kpc, so our jets should be firmly in the self-similar evolutionary phase. This self-similarity is visible from the ribbed structure of the jet spine, which indicates that the outflowing gas is experiencing multiple recollimation shocks (\citealt{vanPutten1996}, \citealt{Bodo2018}, \citealt{Gourgouliatos2018}, \citealt{Smith2019}), which are only present if the jet is not ballistic (\citealt{Falle1991}, \citealt{Bamford2018}). The shocked jet material begins to create a cocoon (lobe) around and ahead of the unshocked jet gas; this is again a feature that occurs only with non-ballistic jets. The inflated lobe expands in a manner similar to a Sedov-Taylor blast wave (\citealt{Sedov1959}).

In the self-similar regime, the shape of the lobes should stay constant with time. From the first few snapshots in Fig. \ref{fig:fig1} we see that the lobes are initially wider at the base than near the head jet. This has to do with the jets exiting the ICM gas density core (at $r\approx25$ kpc) and entering the region of the density profile where $\rho\propto r^{-1.5}$. As the jets leave the core, they experience less resistance and propagate more freely. In the last two snapshots (after the jets have been turned off), we can see the last of the jet gas being shocked. There are also signs of buoyancy beginning to affect the lobes: this is visible as the slight 'break' in the lobes at around half their lengths. The ICM begins to compress and displace the shocked jet gas, initially from the sides, under the action of hydrostatic pressure.

\subsection{Bubble inflation}

\begin{figure*}
\includegraphics[width=0.99\textwidth, trim = 0 10 0 0]{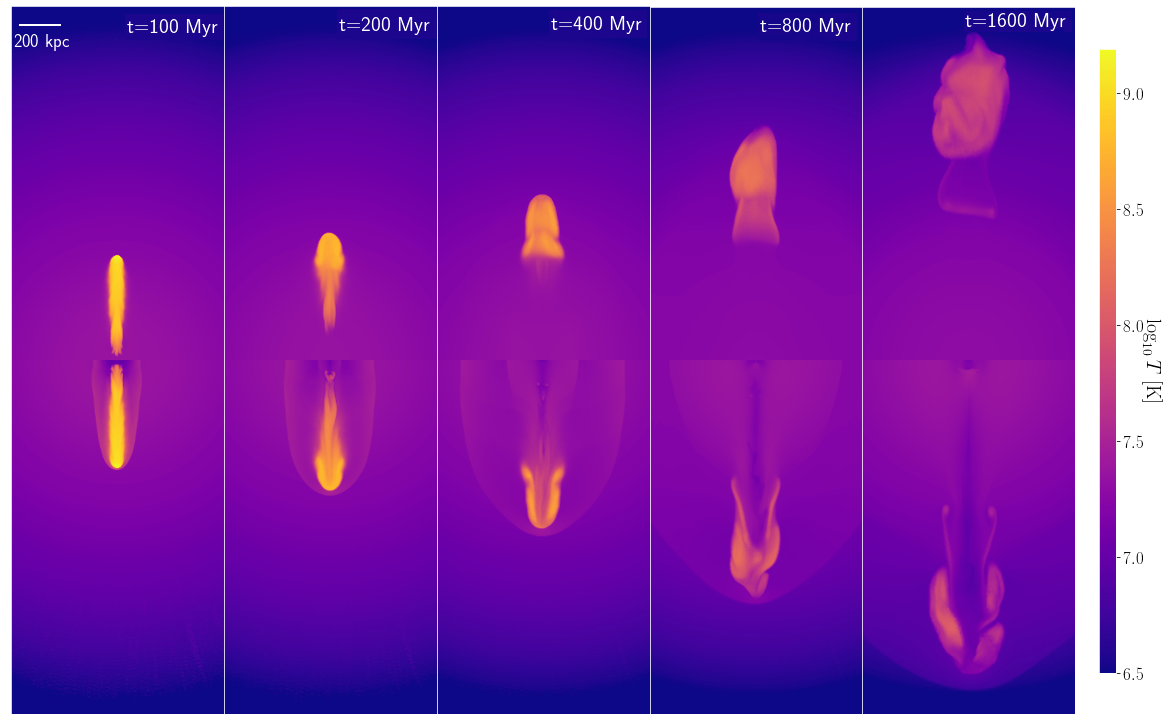}
\caption{Bubble inflation and propagation in our highest resolution simulation ($m_\mr{gas}=10^4$ $\mr{M}_\odot$), at different times. The top half of the simulation is shown through volume renderings of the temperature, in order to highlight the 3D structure of the bubbles. The bottom half shows slices 30 kpc in depth. The panels measure $1200\times360$ kpc$^2$. See caption of Fig. \ref{fig:fig1} for details of the jet launching and the ICM.}
\label{fig:fig2}
\end{figure*}%

In Fig. \ref{fig:fig2} we show the evolution of the shocked jet gas as it transitions from lobes to bubbles. These visualisations are on much larger scales in both time (from twice the jet launching duration, $t=100$ Myr, to $t=1600$ Myr, separated by factors of 2) and physical size (up to $1500$ kpc $\approx1.5$ times the virial radius). The top half of the simulation is shown in terms of volume renderings of the temperature, whereas the bottom half is displayed through temperature slices. We can see that, over time, the hot gas takes the form of a typical mushroom cloud. It also expands and cools on account of adiabatic expansion, but also possibly due to mixing with the ICM. In the first snapshot, the hot gas has a jet-like shape, with the jets having been turned off 50 Myr prior. By the second snapshot, the gas near the jet head has formed a nearly spherical bubble. However, some of the hot gas is still buoyantly rising and joining with the rest of the gas, with that process finishing by the third snapshot. 

In the other snapshots, the hot gas is mostly in the form of nearly-spherical bubbles, but some of that gas is trailing behind the main body of the bubbles. From the temperature slices, we see that this trailing gas takes the form of shedding vortexes. Some asymmetry is visible around the $z-$axis. There are also some asymmetries between the top and bottom bubbles, although they are not obvious here due to the differing ways in which the two halves of the simulation are shown. We have attempted to construct the initial conditions in many ways (see Section \ref{sec:seed_pos}). We find that asymmetries are unavoidable regardless of the initial conditions, even with no seeded perturbations. It is possible that the process by which we stretch our original box to construct a desired density profile introduces perturbations and eliminates perfect symmetry (\citealt{Diehl2015}).

Another feature clearly visible from the slices are filaments of colder gas trailing the bubbles, and connecting them with the centre of the gaseous halo. These filaments are created due to gas uplift that is not unexpected (\citealt{Pope}). We leave the detailed discussion of their properties to Section \ref{sec:drift_wake}.


\subsection{Bubble and ICM energetics}
\label{sec:res_energetics}

We now turn to the question of how much energy is in what form, and in which component. At early times, the different components present in the simulation are the the ambient ICM, the jets (unshocked gas kicked into jets) and the jet-inflated lobes (shocked jet gas). At late times, after the jet is turned off, the jet component begins to disappear as all of the gas is shocked. The lobe of shocked jet gas smoothly transitions into what we refer to as the bubbles. For simplicity, we will not distinguish the lobes and bubbles at these early times, and we will instead refer to them as the bubbles at all times (since we are mostly interested in the late-time evolution in this paper). Note that some of the ICM is entrained into the bubbles; we will not treat this gas as a separate category, and we will attempt to classify this gas as being part of the bubbles. For simplicity, we will also group the jets together with the bubbles. With these choices, we have two categories of gas at all times: the bubbles and the ambient ICM medium.

It is important to consistently determine which particles constitute the bubbles and which ones belong to the ICM. Empirically we have found that the bubble gas, at some given location, is the only gas in the simulation whose density is significantly lower than the expected density at that location (from the initial conditions), while at the same time being significantly hotter. As a result, we define the bubbles as all gas particles whose density $\rho$ and temperature $T$ satisfy $\rho<0.75\rho_0(r)$ and $T>1.25T_0(r)$, respectively, where $\rho_0(r)$ and $T_0(r)$ are the initial density and temperature profiles, and $r$ is the radial distance to the centre of the halo of a gas particle being inspected. Note that this definition includes not only the particles directly kicked into the jets, but also the ICM that is entrained into the lobes of the jet or the bubbles that eventually form from them.

The definition above is somewhat arbitrary, but we find that the particular choices of the numerical factors are not too important. As long as these factors are sufficiently different from 1 (of order $10$ per cent difference, e.g. $0.9$ for $\rho$ and $1.1$ for $T$), other gas that may deviate from its initial density/temperature is not included in this definition. At the same time, most of the gas in the bubbles is much less dense than $\rho_0(r)$ and much hotter than $T_0(r)$. As a result, the choice of the numerical factors does not affect the bubble mass or energetics for the vast majority of their evolution. Instead, it determines how quickly ICM gas being entrained into the bubbles at early times is defined as bubble gas. Similarly, the choice of the numerical factors also determines when the bubble gas that is being mixed with the ICM at late times becomes redefined as part of the ICM. We group the jets together with the bubble gas by including into the bubbles any gas that has a velocity larger than $0.25v_\mr{j}$ (gas that has not yet fully been shocked since being launched into the jets). Any gas that has not been defined as part of the bubbles in this way is defined as part of the ICM. Note that the gas heated by the bow shock is always classified as ICM gas (as it should be), since this gas is both hotter and denser due to being swept up by the jets/lobes/bubbles.

\begin{figure*}
\includegraphics[width=1.01\textwidth, trim = 0 15 0 0]{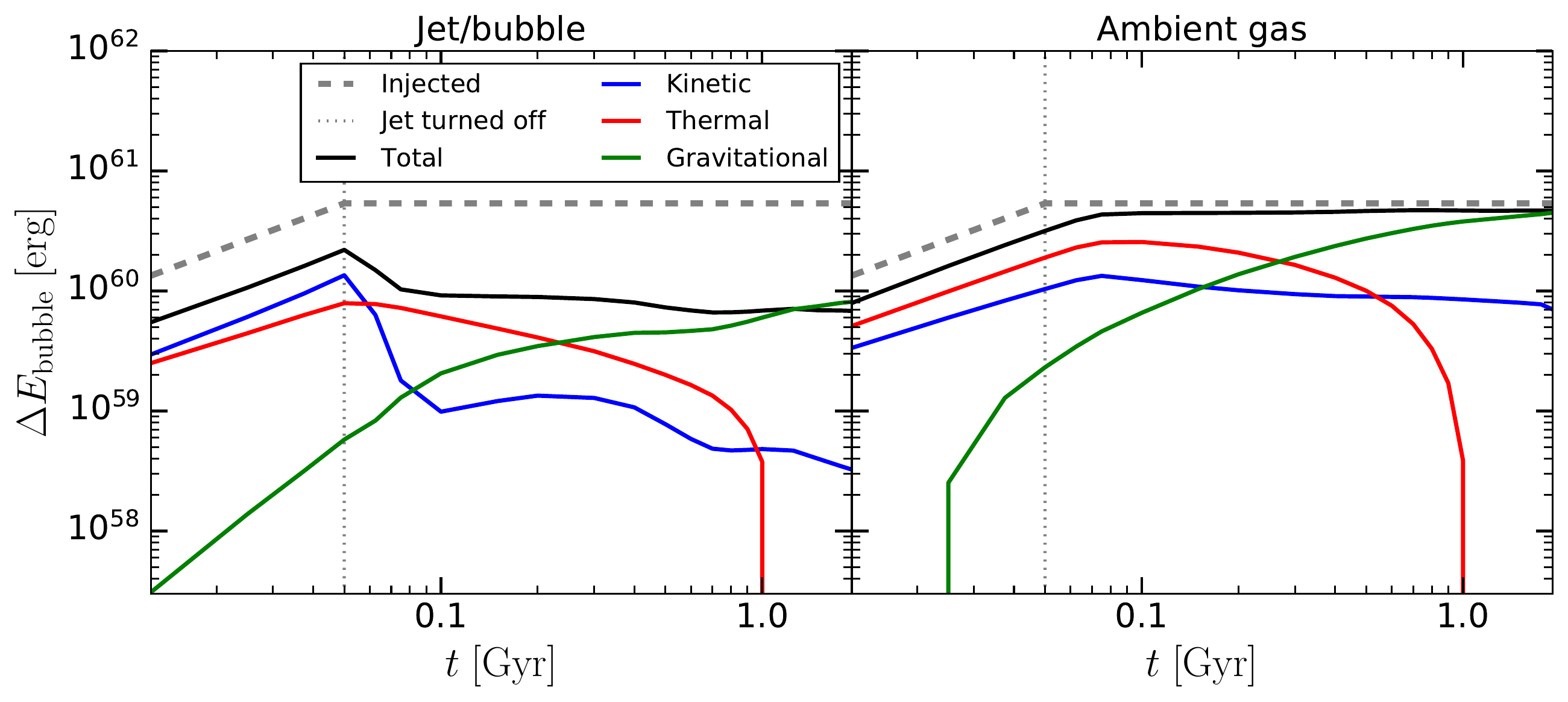}
\caption{Jet/bubble and ambient ICM energies (relative to the energies of the gas in the initial conditions) in our highest resolution simulation ($m_\mr{gas}=10^4$ $\mr{M}_\odot$), at different times. We define all jet particles as part of the bubble, even before the bubble has formed. See caption of Fig. \ref{fig:fig1} for details of the jet launching and the ICM.}
\label{fig:fig3}
\end{figure*}%

We calculate the energies in kinetic, thermal and gravitational potential form for both the bubbles and the ambient ICM. We calculate all of these energies as the total energy of the gas relative to that which the particles had in the initial conditions (i.e. we subtract their initial energies in various forms). These findings are summarized in Fig. \ref{fig:fig3}. We find that during the jet launching phase, most of the energy is efficiently being transferred to the ambient ICM gas (through the bow shock). The fractions of total injected energy in the bubbles versus the ICM are $40$ and $60$ per cent, respectively, and they are constant with time. The bubbles have around $60$ per cent in the kinetic component and $40$ per cent in the thermal component, with the latter dropping slightly with time, likely because the jet is escaping the core and experiencing less shocking while it is being launched. The ambient ICM gas has roughly $30$ per cent in kinetic form and $70$ per cent in thermal. These roughly constant fractions confirm that the jets are in the self-similar regime. The gravitational potential energy is of order a few per cent of the total injected energy while the jets are still on.

As soon as the jets have turned off, the energy in the jets/bubbles begins to drop. This is especially true for the kinetic component, due to almost all of the gas launched into the jets soon being shocked. Despite this kinetic energy being converted into thermal form, the thermal component reaches a peak at around $t=50$ Myr, when the jets are turned off, and begins to drop after that. This is likely due to the lobes of hot gas transferring their energy to the ICM as they are expanding adiabatically. At late times, the relative thermal energy of the bubbles becomes negative, meaning that the bubble gas at these times is colder than in the initial conditions. This is possible since the bubbles have reached the outer regions of the halo, where the ICM temperatures are lower than that in the centre of the halo, and since the typical ratio between bubble and ICM temperature drops with time.

It is also evident from the plot that the bubbles have a roughly constant total energy soon after the jets are turned off, at $10-15$ per cent of the total injected energy. In other words, the bubbles also have to gain energy through some other process. This process is buoyancy, the effects of which can be seen in the gravitational potential energy of the bubbles beginning to dominate after $t=200-300$ Myr. This is supported by visualisations, which show buoyancy clearly beginning to have a strong impact by this time (see Fig. \ref{fig:fig2}). It is also consistent with the bubbles having a roughly constant kinetic energy (due to a constant net upward velocity, which would otherwise fall due to processes such as bow shock launching or drag).

The ambient ICM shares some similarities with the bubbles, in terms of the energy components, after the jets are turned off. The thermal energy quickly reaches a peak and begins to drop. This is possible despite the conversion of kinetic to thermal energy in the bow shock, since the bow shock is also transferring particles to outer regions of the halo (where the temperatures are lower), an effect aided by the buoyant rise of the gas that is mildly heated through interactions with the bubbles. The filaments visible in Fig. \ref{fig:fig2} also play a role, since they are made up of low-entropy gas being uplifted from more central regions of the halo. As they rise, they reduce their temperature in order to come into pressure equilibrium. They have a significant upward velocity, which is visible in the kinetic component of the ICM energy dropping only mildly. At late times, the ICM energy is dominated by the gravitational potential energy. This indicates that the main, long-term consequence of jet feedback is the displacement of particles to larger radii.

\subsection{Effects of jet-inflated bubbles on ICM gas profiles}

\begin{figure*}
\includegraphics[width=1.01\textwidth, trim = 0 20 0 0]{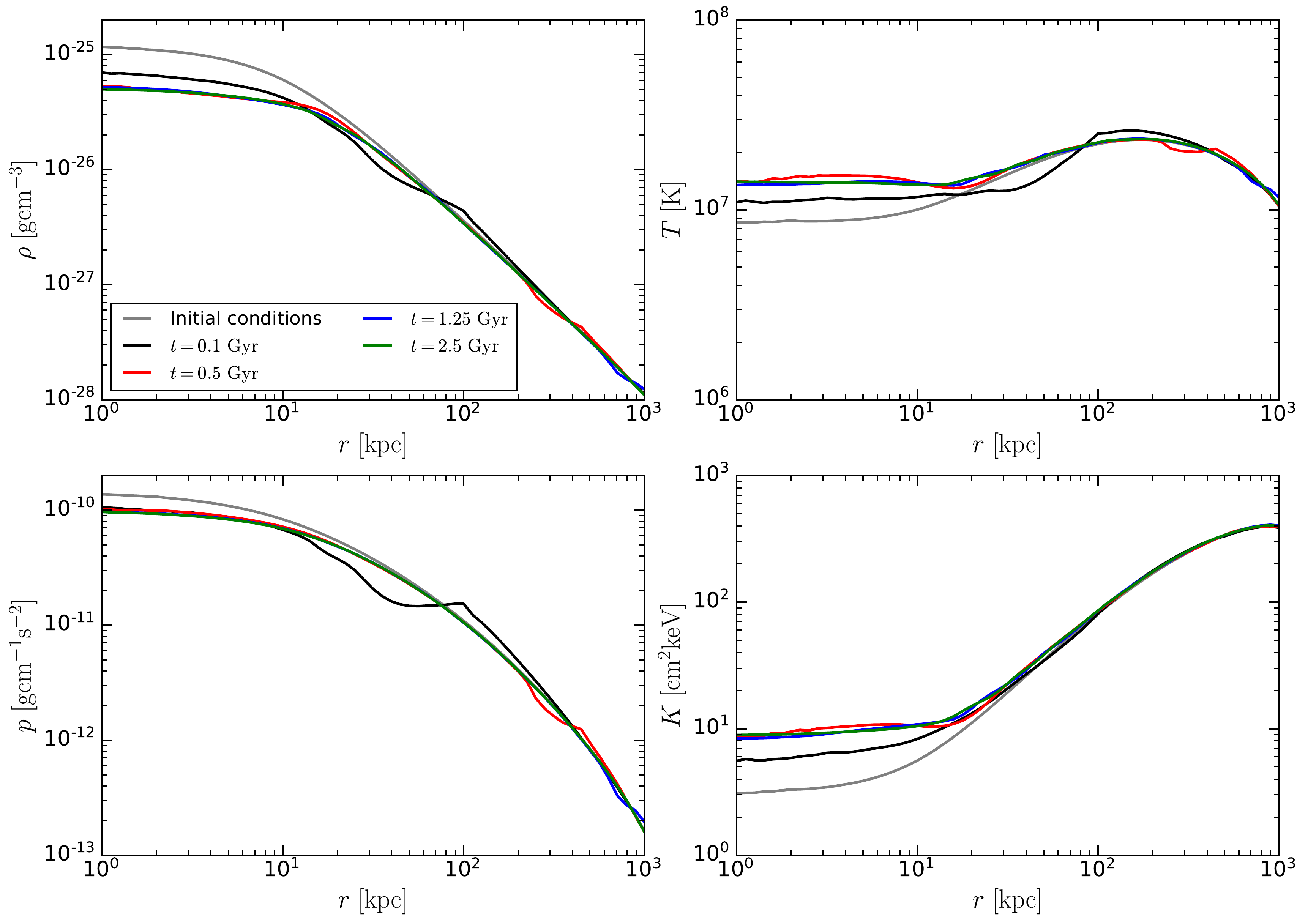}
\caption{Impact of the jets and bubbles on the radial profiles of gas density (top left), temperature (top right), pressure (bottom left) and entropy (bottom right). The profiles are calculated in a mass-weighted fashion, in spherical shells. These results are for our highest resolution simulation ($m_\mr{gas}=10^4$ $\mr{M}_\odot$), at different times as per the legend. The virial radius of the halo is $R_{200}\approx951$ kpc, close to the right-hand end of the $x-$axis. See caption of Fig. \ref{fig:fig1} for details of the jet launching and the ICM.}
\label{fig:fig4}
\end{figure*}%

Additional insight on the effects of jet-inflated bubbles can be gleaned from inspecting their impact on the radial dependence of gas properties, such as density, temperature, thermal pressure and entropy. In Fig. \ref{fig:fig4} we show these profiles at several different times. With the exception of $t=100$ Myr, the gas profiles very closely follow the initial ones at $r>30$ kpc. Within this region, there are significant differences at late times. The core of the new equilibrium profile, after the passage of the jets (bubbles) is around two times less dense, $70$ per cent hotter and has a $30$ per cent lower thermal pressure and $2.5$ times higher entropy. These changes in the core are likely due to a combination of both heating and transfer of gas to larger distances (through either the bow shock or uplift, we discuss the latter in detail in Section \ref{sec:drift_wake}).

In addition to the global changes in the profiles, the effects of the jets are visible as local features. At $t=100$ Myr we can see signs of a high-density, high-temperature, and over-pressured region at $r=100-200$ kpc. Behind this region, between $20$ and $80$ kpc, we find lower-density, lower-temperature and under-pressured gas. This configuration is typical of shock waves, so we interpret this as the bow shock. The bubbles are not visible on these plots due to spherical averaging. At later times, spherical averaging makes even the bow shocks hard to discern.

Our results here indicate that the jets/bubbles have transported significant amounts of material from the inner portions of the gaseous halo out to larger radii (beyond the virial radius at late times). We will now look at similar profile plots, but at one time ($t=400$ Myr), and showing the profiles along the jet/bubble axis, as well as in other directions. First we calculate profiles within a $10\degree$ cone around the $z-$axis (on both sides), ensuring that at any given radius, the properties of the gas are dominated by only one feature (e.g. the jets or bow shocks). We then calculate profiles using all gas particles outside similar cones around the $z-$axis, but with a half-opening angle of $30\degree$. This profile serves as a basis against which we can compare the first one, although it does also include the laterally expanding bow shock.

\begin{figure*}
\includegraphics[width=1.01\textwidth, trim = 0 15 0 0]{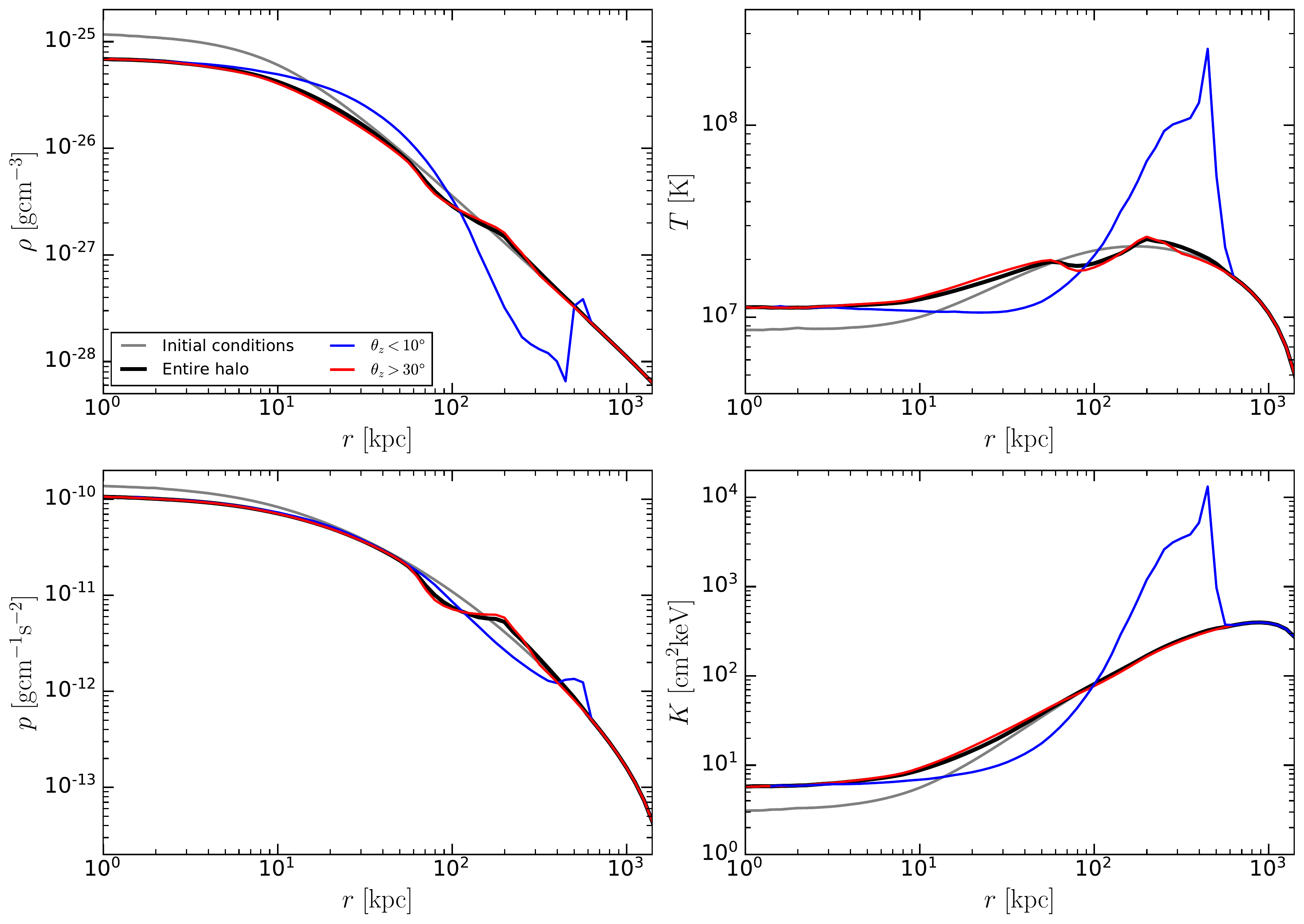}
\caption{Impact of the jets and bubbles on the radial profiles of gas density (top left), temperature (top right), pressure (bottom left) and entropy (bottom right). The profiles are calculated in a mass-weighted fashion. These results are for our highest resolution simulation ($m_\mr{gas}=10^4$ $\mr{M}_\odot$), at $t=800$ Myr. We show the profiles using all particles (black), using particles only within $10\degree$ of the jet-launching axis (blue), and outside $30\degree$ of the jet launching axis (red). The virial radius of the halo is $R_{200}\approx951$ kpc, close to the right-hand end of the $x-$axis. See caption of Fig. \ref{fig:fig1} for details of the jet launching and external medium.}
\label{fig:fig5}
\end{figure*}%

In Fig. \ref{fig:fig5} we show the profiles calculated in the above manner, with an average profile (using all gas particles) also shown. Along the jet-launching axis, we see the effects of the bow shocks at the largest radii ($r=500$ kpc) as high-density, high-temperature, over-pressured gas$-$this is ICM gas that has been swept up into the bow shock. Immediately within the bow shocks is even hotter gas that has a very low density and a very high entropy$-$this gas belongs to the bubbles composed of jet material that has been shocked. They are close to pressure equilibrium with the ICM. At distances even smaller than the bubbles ($r<100$ kpc), we find lower-entropy, lower-temperature and higher-density gas that extends from the centre of the bubbles smoothly down to the centre of the gaseous halo. This is uplifted gas that we discuss in the next section. We note, that this gas has slightly higher temperatures and lower densities than the core of the initial gas profile, presumably due to the core having been heated by the jets through the bow shocks, or due to the uplifted gas originating mainly from larger radii.

Turning now to the profiles outside the cones near the $z-$axis, we find the bow shocks, as well as lower-density and lower-temperature gas immediately following the bow shocks. This is a result of the bow shocks sweeping up gas, so the profiles take some time to settle down to a new equilibrium. We find that this profile is very similar to the average profile (using the whole halo). Note that these features are visible at radii smaller than $r=500$ kpc, the distance to the bow shock along the $z-$axis, due to its ellipsoidal shape.

\section{The uplift of ambient ICM gas behind jet-inflated bubbles}
\label{sec:drift_wake}

As we have seen from some of the results presented in the previous Section, at late times in the simulations, we find that the ambient ICM gas behind the bubbles has a lower entropy, higher density and lower temperature than the ICM gas surrounding it. In fact, this gas even has a lower temperature than its starting temperature in the initial conditions. In this Section we will consider some general properties of the bubbles and ICM gas, focusing mostly on the latter, and including the drift and wake that form behind the bubbles.

\subsection{General properties of gas at late times}

\begin{figure*}
\includegraphics[width=1.01\textwidth, trim = 0 15 0 0]{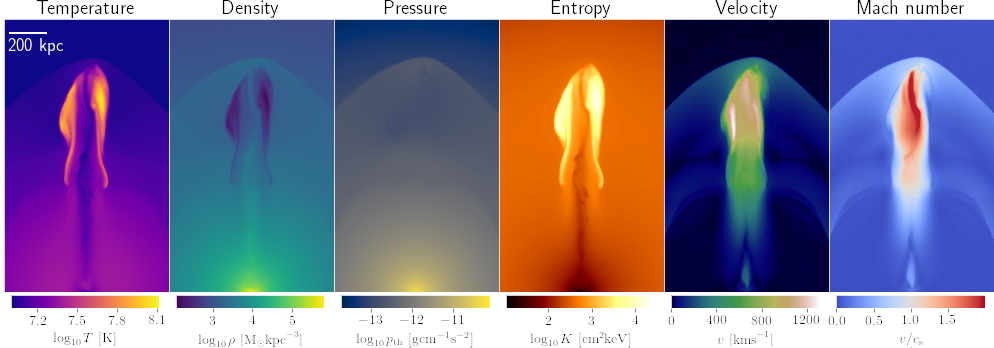}
\caption{Slices through the gas ($30$ kpc depth) $800$ Myr after the start of the simulation, showing the structure of the jet bubbles and their surroundings, at a resolution level of $m_\mr{gas}=10^4$ $\mr{M}_\odot$. The panels measure $1200\times360$ kpc$^2$. Each panel shows a different property, as in the panel titles and coloured as per the colour bars below the panels. See caption of Fig. \ref{fig:fig1} for details of the jet launching and ICM.}
\label{fig:fig6}
\end{figure*}%

In Fig. \ref{fig:fig6} we show visualisations of various gas properties in our highest-resolution simulation at $t=800$ Myr. These panels show slices of $30$ kpc in depth, so projection effects do not hide any features in these plots. Most obviously, we can see the jet-inflated bubbles of hot, low-density and high-entropy gas. The typical temperature of the bubble gas has fallen to no more than $\approx10^{8}$ K by this point, and the density ratio relative to the ICM is of order a factor of $0.1$. The bubbles are in almost perfect pressure equilibrium with their surroundings. They are moving with velocities of order $1000$ kms$^{-1}$, or less, which is similar to the sound speed of the medium. In Fig. \ref{fig:fig6} we can also see the bow shocks moving ahead of the bubbles. The gas being shocked has virial temperatures ($\approx10^{7.5}$ K), is over-dense and over-pressured. It is moving at the sound speed, with a Mach number of $\approx1$ or slightly above. 

In the central regions of the gaseous halo, within $600$ kpc, we find what looks like a core of high temperature and high density gas that is of a higher pressure than the rest of the ICM. This gas makes up the new equilibrium gas profile being established just inside the propagating bubbles and bow shock fronts. This can also be seen on the velocity plots, as this gas has a net zero velocity. Gas just outside this core has a net radial velocity inwards (not visible on this plot). This is gas that was swept outwards by the bow shocks, but is now falling and settling down to a new equilibrium.

The last feature visible in Fig. \ref{fig:fig6}, which we have not yet discussed, are the two filaments of cold ($10^7$ K), somewhat dense and low-entropy gas connecting each bubble to the centre of the halo. These filaments are a result of uplift of ambient, low-entropy gas from central parts of the gaseous halo (\citealt{Pope}), in the form of the drift (\citealt{Darwin1953}, \citealt{Dabiri2006}) and wake (\citealt{Yang2003}, \citealt{Crowe2005}). The former arises due to the displacement of the ICM by the moving bubbles, whereas the latter is a secondary feature of buoyancy. The drift should correspond to the main body of the filaments, while the wake should be cold gas in an indentation in the bubbles. We do indeed find cold, ICM gas in the centre of the bubbles. In fact, this gas is among the fastest moving in the simulation, to the point that it is puncturing/deforming the bubble. Visually, however, it is hard to distinguish the drift and wake.

\begin{figure*}
\includegraphics[width=1.01\textwidth, trim = 0 15 0 0]{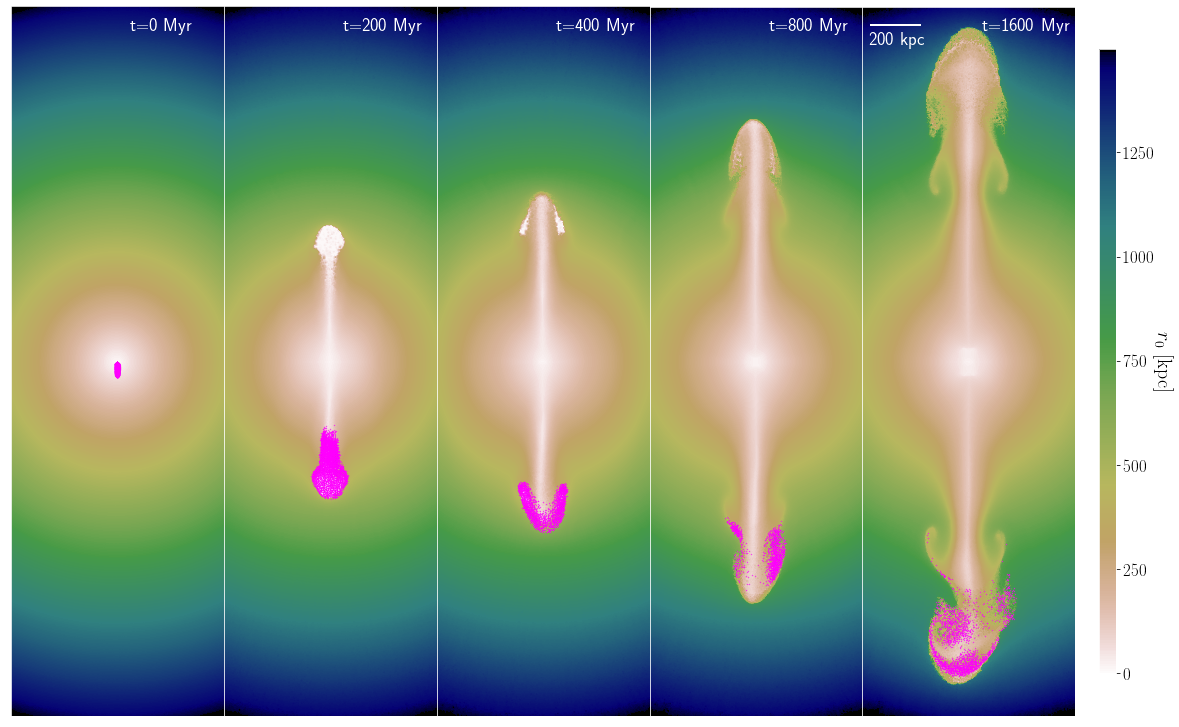}
\caption{Slices ($30$ kpc depth) showing the uplift of gas through the colour-coding of gas particles according to their initial radial distances from the centre of the halo. Gas directly kicked into the jets is not included in the colour map, but we do show individual particles kicked into the jets in the bottom half of the simulation (magenta). See caption of Fig. \ref{fig:fig6} for other details of the simulation.}
\label{fig:fig_uplift}
\end{figure*}

In Fig. \ref{fig:fig_uplift} we show the same gas as in Fig. \ref{fig:fig6} but coloured according to its radii in the initial conditions. In these visualisations we do not include the gas directly kicked into the jets (which also traces the hot gas making up the bubbles). These plots explicitly show that gas is transported from small radii (including the core of the halo) out to large radii. The majority of the uplifted gas originates from within $r<300$ kpc. In the bottom half of the simulation we show the gas particles directly kicked into the jets$-$these particles trace the hot gas making up the bubbles. The visualisations show that the uplifted gas is generally distinct from the hot bubble gas (with the exception of some mixing).

\subsection{The masses and volumes of the drift and wake}

From the discussion in Section \ref{sec:res_energetics}, it is apparent that the filaments, made up from the drift and wake, are energetically important. They are also sufficiently massive to be visually distinct in plots of profiles of various gas properties. Here we will look at the masses and volumes associated with these filaments, separately for the drift and wake. This will also be important for some of our subsequent analysis.

As we have mentioned, the drift and wake are difficult to distinguish in simulations. In a simple theoretical picture, the wake should be any ambient gas within a sphere with the smallest surface area that also encloses the bubbles. We will use a similar definition, but a more general one (applicable for any convex bubble shape). We define the wake as any gas belonging to the ambient medium that resides in the convex hull of the bubble, i.e. any ambient gas within a minimal-area surface that encloses all bubble particles. For this purpose, we use particles originally launched into the jet as tracer particles of the bubble.

The definition of the drift is somewhat more difficult. Visually, from Fig. \ref{fig:fig6}, it is clear that it is low-temperature, high-density and low-entropy gas behind the bubbles. We thus classify a gas particle as part of the drift if it is an ICM particle in a cylindrical region through which the bubbles have traveled, which has an entropy $20$ per cent lower than the initial entropy profile at its location. We find that the exact threshold is relatively unimportant, as long as it is above $10$ per cent.

\begin{figure*}
\includegraphics[width=1.01\textwidth, trim = 0 20 0 0]{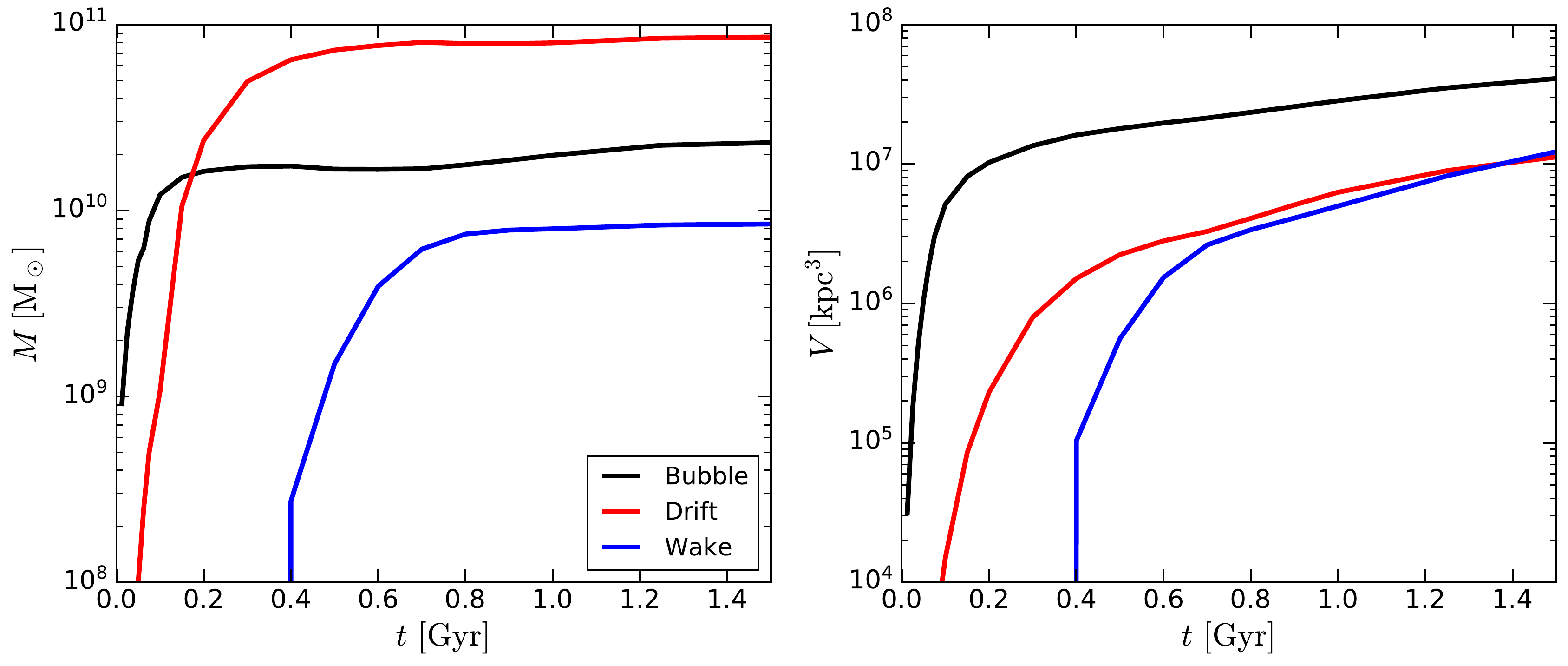}
\caption{Bubble, drift and wake masses and volumes (see main text for definitions and methods of measurement) in our highest resolution simulation, as functions of time. See caption of Fig. \ref{fig:fig1} for details of the jet launching and the ICM.}
\label{fig:fig7}
\end{figure*}%

In Fig. \ref{fig:fig7} we show the masses and volumes of the drift and wake as functions of time, and the same for the bubbles\footnote{We calculate the bubble volume as the sum of $V_i=m_\mathrm{gas}/\rho_i$ over all particles classified as belonging to the bubbles, with $\rho_i$ the density of the $i-$th particle}. as a basis of comparison. The bubbles experience significant entrainment in the first $100$ Myr, which results in an increase in the bubble mass from $M_\mr{b}\approx10^9$ $\mr{M}_\odot$ to $M_\mr{b}\approx1.5 \times 10^{10}$ $\mr{M}_\odot$ (the initially injected mass is $M_\mr{b}=1.35\times 10^9$ $\mr{M}_\odot$). After this initial entrainment, the bubble mass remains nearly constant, with some additional entrainment occurring at late times ($t>800$ Myr), leading to a final bubble mass of $M_\mr{b}\approx2.3 \times 10^{10}$ $\mr{M}_\odot$. The drift begins to appear after $100$ Myr, and it also reaches an approximately constant mass, of around $M_\mr{d}\approx10^{11}$ $\mr{M}_\odot$. The wake appears after $500$ Myr (this is when an indentation in the bubble begins to form, see Fig. \ref{fig:fig2}), and its mass saturates at $M_\mr{w}\approx10^{10}$ $\mr{M}_\odot$. Our results here indicate that drift dominates over wake, which is opposite to the results of \cite{Zhang2022}, who found that the drift is subdominant to turbulent eddy transport, which corresponds to the wake. However, their simulations were 2.5D, were focused on the central regions of the simulated cluster ($r<100$ kpc), and featured bubbles placed by hand (that also held a constant shape and had an imposed velocity), instead of ones that are created by jets. These differences may mean that the results are not directly comparable.

From the plot showing the volumes of each component (right-hand panel of Fig. \ref{fig:fig7}), we see that all of them increase with time, and in a similar way. The bubble volume increases due to adiabatic expansion. The drift volume increases since its volume should be proportional to that of the bubble (see next subsection). The wake volume increases since it is coincident with the cavity in the bubble; if the bubble expands, the cavity will as well.

\subsection{Direct comparison with an analytical model}

We will now compare the masses and volumes of the drift and wake with the model presented in \cite{Pope}. At a minimum, their model posits that bubbles inflated by jets rise through the ICM on account of buoyancy, while also expanding adiabatically, and remaining in pressure equilibrium with the ICM. As we have found, these assumptions are justified. \cite{Pope} also include the effects of drag and mixing with the ambient ICM (entrainment). The drag is not directly measurable from our simulations (since there are many other processes at play), but the entrainment is, as we will show.

The volume of the drift can be related to that of the bubble by a numerical coefficient: $V_\mr{d}=kV_\mr{b}$. This relation can be shown to be true quite generally, in simple models of a sphere moving through a medium at a constant velocity (\citealt{Darwin1953}, \citealt{Benjamin108}, \citealt{Dabiri2006}), although $k$ depends on the shape of the bubble. The effects of the wake can be quantified as additional mass added to the bubble: $M_\mr{w}=qM_\mr{dis,0}$, where $q<1$ is a numerical coefficient and $M_\mr{dis,0}$ is the mass initially displaced by the bubble. According to the entrainment hypothesis (\citealt{Morton1956}), the rate at which ambient material is entrained by the bubble is given by $\dot{M}_\mr{b}=\alpha \rho_\mr{ICM} v_\mr{b} S_\mr{b}$, where $\rho_\mr{ICM}$ is the ambient ICM density, $v_\mr{b}$ the bubble velocity, $S_\mr{b}$ the surface area of the bubble and $\alpha$ is a numerical coefficient that depends on the type of mixing. We refer the reader to \cite{Pope} for a more comprehensive discussion of all three of these effects. 

The values of the numerical coefficients described above have been well determined at least for spherical bubbles, either experimentally or theoretically. In this case, we expect $q=0.24$ based on experiments (\citealt{Yang2003}, \citealt{Crowe2005}) and $k=0.5$ based on theory (\citealt{Darwin1953}). $\alpha$ is usually determined experimentally. It should be around $0.05$ where mixing is due to turbulence in a momentum-driven (jet-like) flow (\citealt{Turner1986}). Jets in the presence of buoyancy have $\alpha=0.065-0.08$, while buoyantly-rising plumes should have $\alpha=0.1-0.16$ (see \citealt{Carazzo2006} for a review). Magnetic fields and viscosity can reduce the value of $\alpha$; the former is not included in our simulations, but the latter is (see \citealt{Borrow2022} for a discussion of artificial viscosity in SPHENIX).

The values of $k$, $q$ and $\alpha$ are unlikely to be truly constant, since our simulation includes many processes and a relatively complex set-up. An additional problem is that the shapes of our bubbles change continuously throughout the simulation. Furthermore, derivations of $k$ for the drift are usually done in the absence of gravity. It is not known what is the interplay between the three effects, and if there is any at all.

Despite the expected difficulties, we will attempt to measure these coefficients. We do so in the following way.
\begin{itemize}
    \item Drift: $k=V_\mr{d}/V_\mr{b}$ (see previous subsection for the definition of the drift and beginning of Section \ref{sec:res_energetics} for the definition of the bubbles).
    \item Wake: $q=M_\mr{w}/M_\mr{dis,0}$ (see previous subsection for the definition of the wake). The initially displaced mass $M_\mr{dis,0}$ can be calculated as the product of the ambient ICM density and bubble volume, $\rho_\mr{ICM} V_\mr{b}$. While there is an arbitrary choice in when this product is taken (since the bubble is not initially placed, but rather forms from a particle distribution that was launched with some velocity), we find that the product is roughly constant with time. We take the value at $t=100$ Myr.
    \item Entrainment: $\alpha=\dot{M}_\mr{b}/ S_\mr{b}\rho_\mr{ICM} v_\mr{b}$. We measure $\dot{M}_\mr{b}$ directly as the derivative in the sum of all bubble particle masses. The surface area of the bubble $S_\mr{b}$ is given by the convex hull that describes the bubble particle distribution, while $v_\mr{b}$ is the average velocity in the z-direction of the bubble particles.
\end{itemize}

\begin{figure}
\includegraphics[width=1.01\columnwidth, trim = 0 20 0 0]{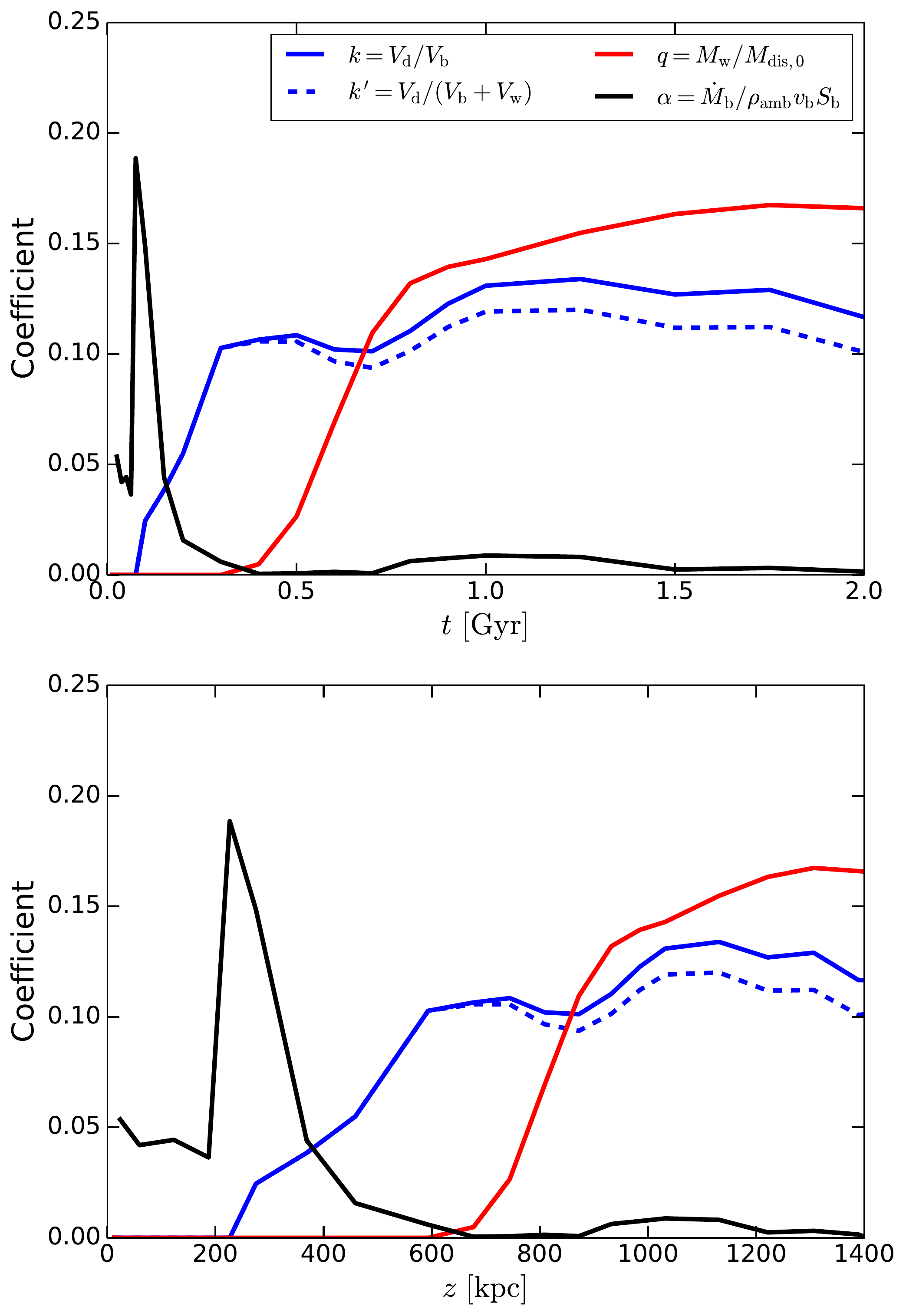}
\caption{Measurement of the drift coefficient $k$, wake coefficient $q$ and entrainment coefficient $\alpha$ (see main text for definitions and methods of measurement) in our highest resolution simulation, as functions of time (top) and bubble height (bottom). We also measure $k^\prime=V_\mathrm{d}/(V_\mathrm{b}+V_\mathrm{w})=k/(1+V_\mathrm{w}/V_\mathrm{b})$, which is the drift coefficient as defined by \protect\cite{Pope}. See text or caption of Fig. \ref{fig:fig1} for details of the jet launching and ICM.}
\label{fig:figm4}
\end{figure}%

In Fig. \ref{fig:figm4} we show the dependence of the three coefficients on time and bubble height in our highest-resolution simulation. The earliest of the three effects to appear in the simulation is entrainment. While the jet is active ($t<50$ Myr), the entrainment coefficient has a value of $\alpha=0.04-0.05$, which is consistent with a momentum-driven flow (\citealt{Turner1986}, \citealt{Dellino2014}), despite the fact that buoyancy is present in this phase, and that we might expect $\alpha\approx0.07$ (\citealt{Carazzo2006}). This is perhaps due to buoyancy beginning to operate only on longer time-scales (several hundred Myr rather than 50 Myr). After the jet is turned off, the entrainment increases to a peak of $\alpha=0.19$, and quickly drops to very small values by $t=300$ Myr ($z=600$ kpc). The peak value is similar to that expected for a buoyantly-rising plume ($\alpha=0.16$; \citealt{Carazzo2006}, \citealt{Suzuki2010}). At late times, the entrainment is very small, usually $\alpha<0.01$. This is potentially due to a lack of resolved small-scale turbulence, which could be either due to the artificial viscosity in SWIFT or an effect of insufficient numerical resolution. 

The drift coefficient grows to $k=0.1$ within $300$ Myr ($z=600$ kpc), and has an approximately constant value ($k=0.1-0.13$) for the rest of the simulation. This is much lower than expected for spherical bubbles in a constant-density medium ($k=0.5$; \citealt{Darwin1953}). The value that we have measured matches our expectation based on the thin nature of the filaments, as is visible in e.g. Fig. \ref{fig:fig6}. The drift in a constant-density medium has a much wider base (see Fig. 1 in \citealt{Pope} for a schematic), but this is likely impossible in a gaseous atmosphere in hydrostatic equilibrium. It is likely that hydrostatic pressure compresses the drift filament, such that it is thinner at the base than near the bubble. In Fig. \ref{fig:figm4} we also show the evolution of the drift coefficient defined in a slightly different way: $k^\prime = V_\mathrm{d}/(V_\mathrm{b}+V_\mathrm{w})$, instead of $k = V_\mathrm{d}/V_\mathrm{b}$. The former is the definition used in \cite{Pope}, whereas we use the latter, which is the more standard definition (although one that may be less physically motivated in the presence of the wake).

The wake coefficient grows to an approximately constant value of $q=0.15-0.17$ by $t=800$ Myr ($z=1000$ kpc), but is negligible at $t<500$ Myr ($z<700$ kpc). This value is lower than expected for a spherical bubble ($q=0.24$; \citealt{Yang2003}, \citealt{Crowe2005}), for a few possible reasons: i) the bubbles form from gas that is initially moving, ii) the bubbles are typically somewhat elongated, iii) the gravitational field weakens as the bubble rises. 

\subsection{Inferring drift, wake and entrainment from an analytical model for the evolution of the bubble velocity}

\begin{figure*}
\includegraphics[width=1.01\textwidth, trim = 0 15 0 0]{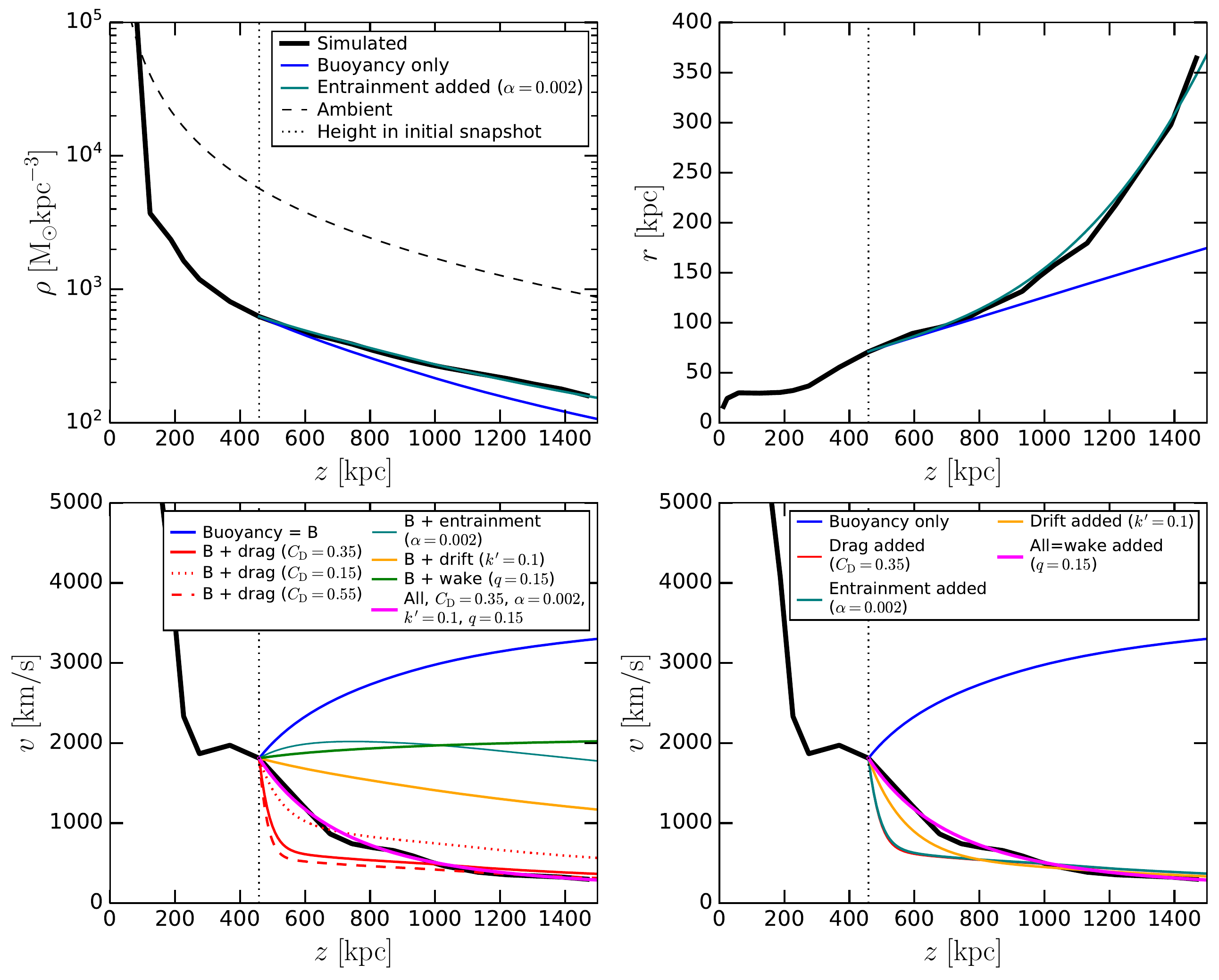}
\caption{Bubble properties measured from our highest resolution simulation (see main text for methodology), as functions of bubble height (black lines). The panels show density (top left), radius (top right), and velocity (bottom). These are compared with predicted values from the analytical model by \protect\cite{Pope}. The initial conditions for this model are our measured values at $100$ Myr, corresponding to a bubble height of $z=450$ kpc. In the bottom left panel, different lines represent model predictions with different effects included; the blue line represents buoyancy and adiabatic expansion, while all other lines, with the exception of magenta, model one additional effect (as per the legend). In the bottom right panel, lines of those same colours represent each effect included successively. Magenta lines represent model predictions with all effects included. Numerical coefficients for each effect are given in the legends. In the top two panels we do not show model predictions with drag, entrainment, drift or the wake, as they do not influence bubble densities or radii. See caption of Fig. \ref{fig:fig1} for details of the jet launching and ICM.}
\label{fig:fig8}
\end{figure*}%

\cite{Pope} present an analytical model for the evolution of a rising bubble that includes buoyancy, adiabatic expansion, drag, drift, wake and entrainment. They quantified these effects through the numerical coefficients described above and developed a system of differential equations that can be used to evolve the bubble. The drift and wake serve as additional terms in the inertia of the bubbles, whereas entrainment enters the equations as a mass flux term. The only forces acting on the bubble are buoyancy and drag. The buoyancy force is equal to $F_\mathrm{B}=(\rho_\mathrm{ICM}-\rho_\mathrm{b})gV_\mathrm{b}$, where $\rho_\mathrm{ICM}$ and $\rho_\mathrm{b}$ are the ICM and bubble densities, respectively, $g$ is the gravitational acceleration and $V_\mr{b}$ the bubble volume. The drag force is modeled using the usual formula: $F_\mr{D}=-(1/2)C_\mr{D}\rho_\mr{ICM} A_\mr{b} v_\mr{b}^2 $, where $C_\mr{D}$ is the drag coefficient (dependant on the shape of the bubble and the Reynolds number), $v_\mr{b}$ the bubble velocity and $A_\mr{b}$ its cross-sectional area.

The system of equations developed by \cite{Pope} has three dependant variables: the density $\rho_\mr{b}$, radius $r_\mr{b}$ and velocity $v_\mr{b}$ of the bubble. The independent variable is the radial distance (height) of the bubble $z$. The system of differential equations can be written as:
\begin{equation}
\frac{\mathrm{d} r}{\mathrm{~d} z}=\alpha -\frac{\mu m_{\mathrm{P}} r g}{5 k_{\mathrm{B}} T_{\mathrm{ICM}}},
\label{eq:Pope1}
\end{equation}
\begin{equation}
\frac{\mathrm{d} \rho_{\mathrm{b}}}{\mathrm{d} z}=\frac{3 \alpha}{r}\big(\rho_{\mathrm{ICM}}- \rho_{\mathrm{b}}\big)+\frac{5\mu m_{\mathrm{p}} \rho_{\mathrm{b}} g}{3 k_{\mathrm{B}} T_{\mathrm{ICM}}},
\label{eq:Pope2}
\end{equation}
\begin{equation}
\begin{aligned}
\frac{\mathrm{d} v_\mr{b}}{\mathrm{~d} z}= &\frac{-6 \alpha v_\mr{b} \rho_{\mathrm{ICM}} / r}{\rho_{\mathrm{tot}}} + \frac{g}{v_\mr{b}} \frac{\left(\rho_{\mathrm{tot}}-(1+k^\prime) \rho_{\mathrm{ICM}}\right)}{\rho_{\mathrm{tot}}} \\
&-\frac{2 \mu m_{\mathrm{p}} v_\mr{b} g}{5 k_{\mathrm{B}} T_{\mathrm{ICM}}} \frac{k^\prime \rho_{\mathrm{ICM}}}{\rho_{\mathrm{tot}}} -\frac{F_\mr{D}}{V_{\mathrm{b}} v_\mr{b} \rho_{\mathrm{tot}}},
\end{aligned}
\label{eq:Pope3}
\end{equation}
where $T_\mr{ICM}$ is the ICM temperature and $\rho_{\mathrm{tot}}=\rho_\mr{b}+k^\prime\rho_\mr{ICM}+qM_\mr{dis,0}/V_\mr{b}$ an effective total density, with $qM_\mr{dis,0}$ the wake mass. Here, $k^\prime$ is the drift coefficient as defined in \cite{Pope}, which relates to our definition through $k^\prime=k/(1+V_\mathrm{w}/V_\mathrm{b})$. The gravitational field $g$ and the densities and temperatures of the ICM, $\rho_\mr{ICM}$ and $T_\mr{ICM}$, respectively, all depend on the height of the bubble (as determined by the radial profiles of gas quantities calculated under the assumption of hydrostatic equilibrium). Note that we have written equations (\ref{eq:Pope1}-\ref{eq:Pope3}) by already assuming that both the ICM and the bubble have an adiabatic index $\gamma=5/3$, since we assume the same for our simulations (in reality the bubble would have an adiabatic index of $4/3$, see \citealt{Pope} for a more general system of equations with arbitrary $\gamma_\mr{ICM}$ and $\gamma_\mr{b}$). In addition, the bubble is assumed to be spherical ($V_\mr{b}=4\pi r^3\rho_\mr{b}/3$ and $A_\mr{b}=4\pi r^2$).

Equations (\ref{eq:Pope1}-\ref{eq:Pope2}) can be derived from considering entrainment (first terms in both equations) and adiabatic expansion (second terms in both equations). Note that in the absence of entrainment ($\alpha=0$), the evolution of bubble radii and densities with height are decoupled from each other. When entrainment is added ($\alpha>0)$, the density falls less quickly with height due to the presence of the positive first term in equation (\ref{eq:Pope2}). The bubble radius also increases more quickly with height. Equation (\ref{eq:Pope3}) is the momentum equation, with the first term corresponding to entrainment, the second term to buoyancy, the third term to adiabatic expansion and the last term to drag.

Without the effects of drag, the wake, drift or entrainment, the adiabatic expansion of the bubbles leads to a slower drop in density than that of the ICM (that falls as $1/r^{1.5}$ in our case), meaning that the bubble will stop accelerating due to buoyancy at some distance where its density becomes equal to that of the ICM. The inclusion of drag leads to the bubble rising at a terminal velocity (that can weakly depend on radius, since the gravitational acceleration does as well). Including drift, wake or entrainment will make the bubbles harder to either accelerate or decelerate (depending on whether buoyancy or drag dominates).

In order to solve the system of equations (\ref{eq:Pope1}-\ref{eq:Pope3}), one needs to assume some values for the coefficients $C_\mr{D}$, $k^\prime$, $q$, and $\alpha$. The initial conditions for the system are the bubble radius, density and velocity at some height. In order to compare our simulations with the model developed by \cite{Pope}, we choose the initial conditions for the system of equations (\ref{eq:Pope1}-\ref{eq:Pope3}) from our own simulations at 100 Myr, since this is the first snapshot by which all of the gas launched into the jets has been shocked. The shocked jet gas (i.e. the jet lobes) are still very elongated, as they have not yet been reshaped by buoyancy into a more bubble-like shape.

We measure the bubble properties (height, radius, density and velocity) at every snapshot in the following way. We use the particles originally launched into the jet as tracer particles, since we find that they trace the bubble shape very well. These particles are not the only ones that constitute the bubbles (due to entrainment, especially at higher resolutions), but provided that mixing is strong enough, the properties of these particles should trace those of the overall bubble on average. We first calculate their centre of mass (a point along the $z-$axis). We define the height of the bubble as the distance of the centre of mass from the origin. Its radius is defined as the average cylindrical radius of the 20 particles farthest from the $z-$axis. The bubble velocities and densities are calculated as the mean values using all particles classified as making up the bubbles (using the same definition of the bubbles as in Section \ref{sec:res_energetics}). The height of the bubbles in the initial conditions that we use for our analysis here (i.e. the simulation snapshot at $t=100$ Myr) is $z=450$ kpc. 

In Fig. \ref{fig:fig8} we show these properties of the bubbles in our highest-resolution simulation as a function of bubble height. The drag, drift and wake do not play a role in the evolution of bubble density and radius in this model, which are shown in the top two panels. For this reason, we can use our simulated bubble densities and radii to infer how much entrainment is occurring (regardless of the bubble velocities). The decrease in the bubble density with height is consistent with entrainment being present at these late times, although with a very small value of the entrainment coefficient $\alpha$. The evolution is consistent with the entrainment coefficient being equal to $\alpha=0.002$, with no entrainment leading to somewhat lower densities. The evolution of bubble radii is consistent with the same value of $\alpha$. This value is similar to the value that we directly measured (Fig. \ref{fig:figm4}).

In the bottom left panel of Fig. \ref{fig:fig8}, we show the bubble velocity as a function of bubble height. The bubble velocity in the simulation drops sharply by $250$ kpc, as the outflowing jet material is strongly shocked. It reaches a small local minimum, and then begins to rise on account of buoyancy. It reaches a local maximum velocity of $2000$ kms$^{-1}$ at $400$ kpc, and then drops slowly, beginning to asymptote to a value of $150$ kms$^{-1}$ at large distances. We compare this velocity evolution with that predicted by the \cite{Pope} model by including the relevant effects one by one. As already mentioned, the initial conditions used for the model are the measured values shown on these plots, at 100 Myr (bubble height of $450$ kpc). 

As can be seen in the bottom left panel of Fig. \ref{fig:fig8}, including only buoyancy and adiabatic expansion in this comparison leads to an increase in the velocity (but also a saturation, which is not visible in this plot since it occurs at larger heights), due to adiabatic expansion leading to the bubble becoming comparable in density to the ICM). We then add drag, and show several curves with different drag coefficients (a moderate value of $C_\mr{D}=0.35$, close to the typical value for bullet-shaped objects at large Reynolds numbers ($C_\mr{D}=0.3$), as well as a low value, $C_\mr{D}=0.15$, and a high value, $C_\mr{D}=0.55$). Adding drag changes the picture considerably. The initial deceleration of the bubble is consistent with the smallest value of the drag coefficient shown ($C_\mr{D}=0.15$), and may even warrant a value smaller than that. Note that in this portion of the bubble's evolution ($z<700$ kpc), the drift, wake and entrainment are roughly negligible in our simulations (especially the latter two), as measured and shown in the bottom panel of Fig. \ref{fig:figm4}). This means that the inferred value of the drag coefficient is indicative of the actual behaviour of the shocked jet gas (which is transitioning into a bubble), and can be considered an indirect measurement of this parameter.

At intermediate times (bubble heights), the behaviour of the velocity cannot be used to infer the drag coefficient in isolation, largely due to the presence of the drift and wake. However, at very late times (i.e. bubble heights of $z>1000$ kpc), the effects of the drift and wake become negligible in the evolution of the bubble velocity (i.e. the first and third term in Eqn. \ref{eq:Pope3} become negligible compared to the second and last term). The remaining effects that have an impact on the bubble velocity are buoyancy and drag. These two forces balance each other, leading to a terminal velocity (that may weakly depend on height). This is the velocity that we measure at late times in our simulation. The two higher values of the drag coefficient shown in the bottom left panel of Fig. \ref{fig:fig8}, $C_\mr{D}=0.35$ and $C_\mr{D}=0.55$, are both in fairly good agreement with the measured velocity evolution, with the latter comparing slightly better. This value is close to the drag coefficient of a spherical bubble, which is close to the actual bubble shape (see last panel of Fig. \ref{fig:fig2}). Due to other effects being negligible, this can be considered an indirect measurement of the drag coefficient at these late times. This drag coefficient is larger than at early times, indicating significant time evolution of the drag coefficient.

In the bottom left panel of Fig. \ref{fig:fig8} we also show the evolution of the bubble velocity if other effects (the drift, wake and entrainment) are included alongside buoyancy and adiabatic expansion (but one by one and separate from each other, and also separate from drag). For the drift and wake we test values of $k^\prime=0.1$ and $q=0.15$, respectively, roughly matching the values we measured (in Fig. \ref{fig:figm4}). Modeling only the drift leads to a decline in the velocity, but much weaker than caused by drag. Modeling only the wake leads to an increase and subsequent saturation of the velocity. Including only entrainment (with $\alpha=0.002$, as inferred from the evolution of the bubble density and radii) alongside buoyancy and adiabatic expansion leads to a peak and subsequent decline. 

Finally, with the magenta line we show the evolution of the bubble velocity if all of these processes are included simultaneously. The coefficients chosen for this evolution are the same as discussed above, with the drag represented using a moderate value of the drag coefficient, $C_\mr{D}=0.35$. The evolution using the \cite{Pope} model and these coefficients closely matches the simulation. As we have already discussed, and as confirmed by the discussion in the above few paragraphs, this is an oversimplification. In particular, the drag coefficient likely depends on time (bubble height), and the drift and wake coefficients become appreciable only at later times (later than the initial snapshot when we begin the evolution using the \citealt{Pope} model). We have decided against incorporating time-dependant coefficients or turning on the drift and wake at later times, since various choices can be made that also lead to the correct evolution of the bubble velocity. We have instead chosen the simple version of the model (with constant coefficients). It is likely that in the analytical model, at early times, the large value of the drag coefficient (too large compared to the simulation) is compensated by the presence of the drift and wake in the model at these times (effects which are not yet present in the simulation). The drift and wake work to slow down the deceleration due to drag (see below), probably mimicking the effect of choosing a lower drag coefficient. In our simulated bubbles, however, it is likely that the effects of the drift, wake and entrainment are negligible, and the behaviour is dominated by a time-dependent drag coefficient.

In the bottom right panel of Fig. \ref{fig:fig8} we show the same simulated velocity evolution as in the bottom left panel, but the lines showing the predicted evolution now include each effect successively. We start off with only buoyancy and adiabatic expansion, and then add drag, entrainment, drift and wake, in that order. We use the same coefficients as discussed above. Including entrainment on top of drag leads to a slightly quicker decline in the velocity, as compared to using only drag. Adding drift causes a slower decline in the velocity. Similarly, adding the wake on top of that makes the velocity drop even more slowly, yielding the final prediction, which agrees well with our measured evolution of bubble velocity. The last two changes may appear somewhat counter-intuitive: why would adding the drift or wake make the bubbles decelerate less quickly, considering the fact that they individually decelerate the bubbles in isolation? The reason they delay the deceleration is due to the drift and wake being modeled as added mass (inertia) of the bubbles$-$this is not immediately obvious from Eqn. (\ref{eq:Pope3}) but it is easy to see in the original form of the equation as written in \cite{Pope} (this equation is essentially Newton's second law of motion). This means that the bubbles are harder to both accelerate and decelerate. In a situation where deceleration dominates (due to drag overcoming buoyancy), the bubbles will be decelerated more slowly, as the drift and wake must also decelerate in tandem with the bubbles.

\begin{figure}
\includegraphics[width=1.01\columnwidth, trim = 0 15 0 0]{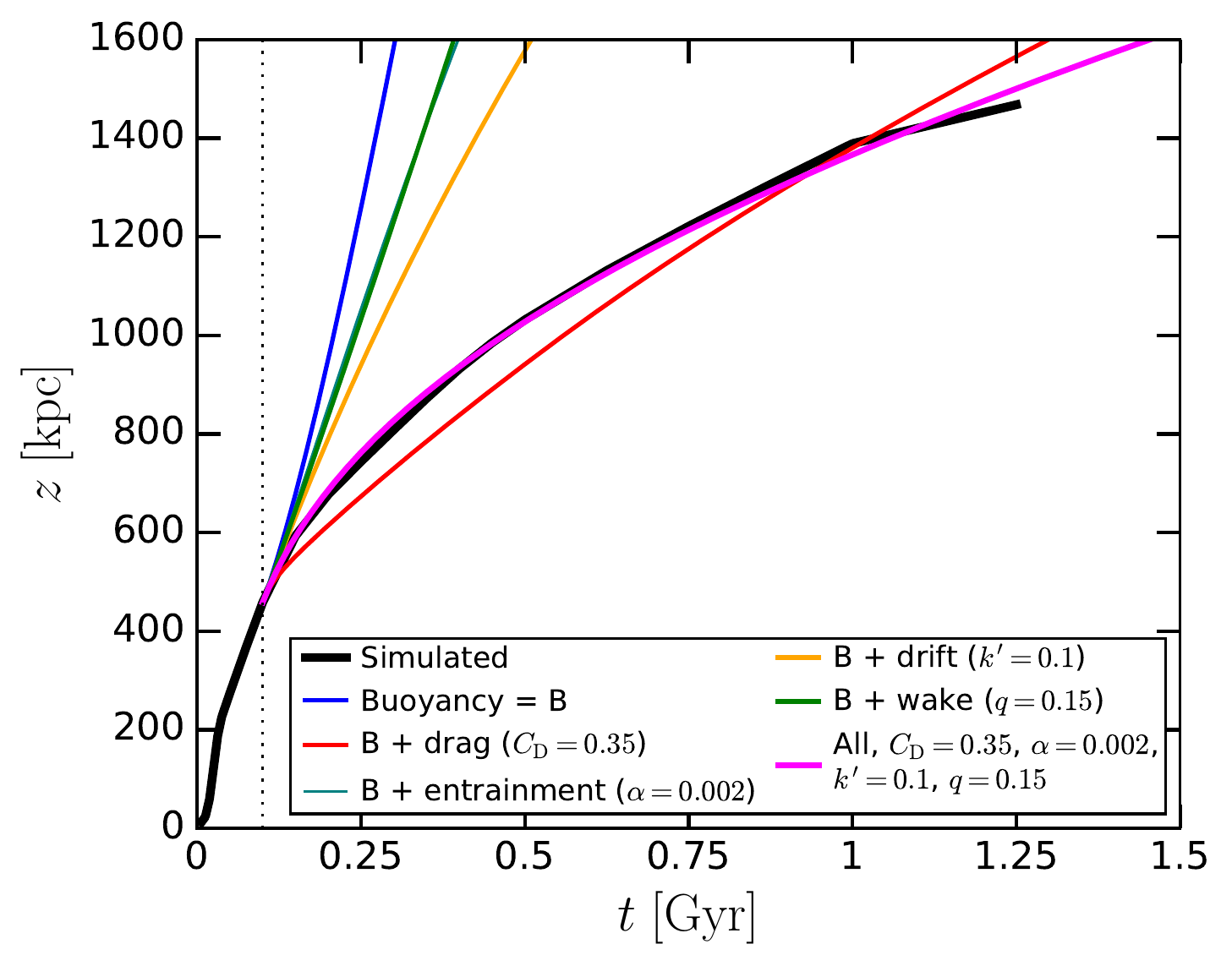}
\caption{Evolution of the simulated bubble height with time, compared with predicted values from the analytical model by \protect\cite{Pope}. The details of the simulation and analytical evolution are the same as in Fig. \ref{fig:fig8}. Different lines represent model predictions with different effects included one by one (same as bottom left panel of Fig. \ref{fig:fig8}). The magenta line represents model predictions with all effects included.}
\label{fig:figm4_z_vs_t}
\end{figure}%

In Fig. \ref{fig:figm4_z_vs_t} we show the dependence of the simulated bubble height on time, compared to the predictions using the \cite{Pope} model. Here we show the same set of predictions as in the bottom left panel of Fig. \ref{fig:fig8}. The predicted evolution of the height with time corresponds simply to an integral of the velocities presented in the same figure. The measured values of the bubble height, however, are obtained from the simulation independently from the velocities. This means that the comparison is somewhat independent from the comparison of the velocity evolution. We find that the magenta line, representing the full evolution with all effects included (and the same coefficients as in Fig. \ref{fig:fig8}), tracks the simulated values well.

\section{Jet-inflated bubbles with varying parameters}
\label{sec:param_study}

In the previous sections we focused on general features of jet-inflated bubbles, as well as features of the ICM that arise on account of these bubbles. These analyses were done for our highest resolution simulation ($m_\mr{gas}=10^4$ $\mr{M}_\odot$), which has a jet power $P_\mr{j}=3.16\times10^{45}$ ergs$^{-1}$, launching velocity $v_\mr{j}=2\times10^4$ kms$^{-1}$ and opening angle $\theta_\mr{j}=15\degree$. Here we will broaden our analysis by varying all of these parameters. In addition, we vary the hydrodynamical scheme used in the simulations. The standard resolution at which we vary other parameters is $m_\mr{gas}=10^5$ $\mr{M}_\odot$, and our standard hydrodynamical scheme is SPHENIX. The standard set of physical parameters is the same as listed above for the high-resolution simulation. We keep the jet duration the same in all cases ($t=50$ Myr), and we also keep the ICM unchanged.

\begin{figure*}
\includegraphics[width=0.81\textwidth, trim = 0 15 0 0]{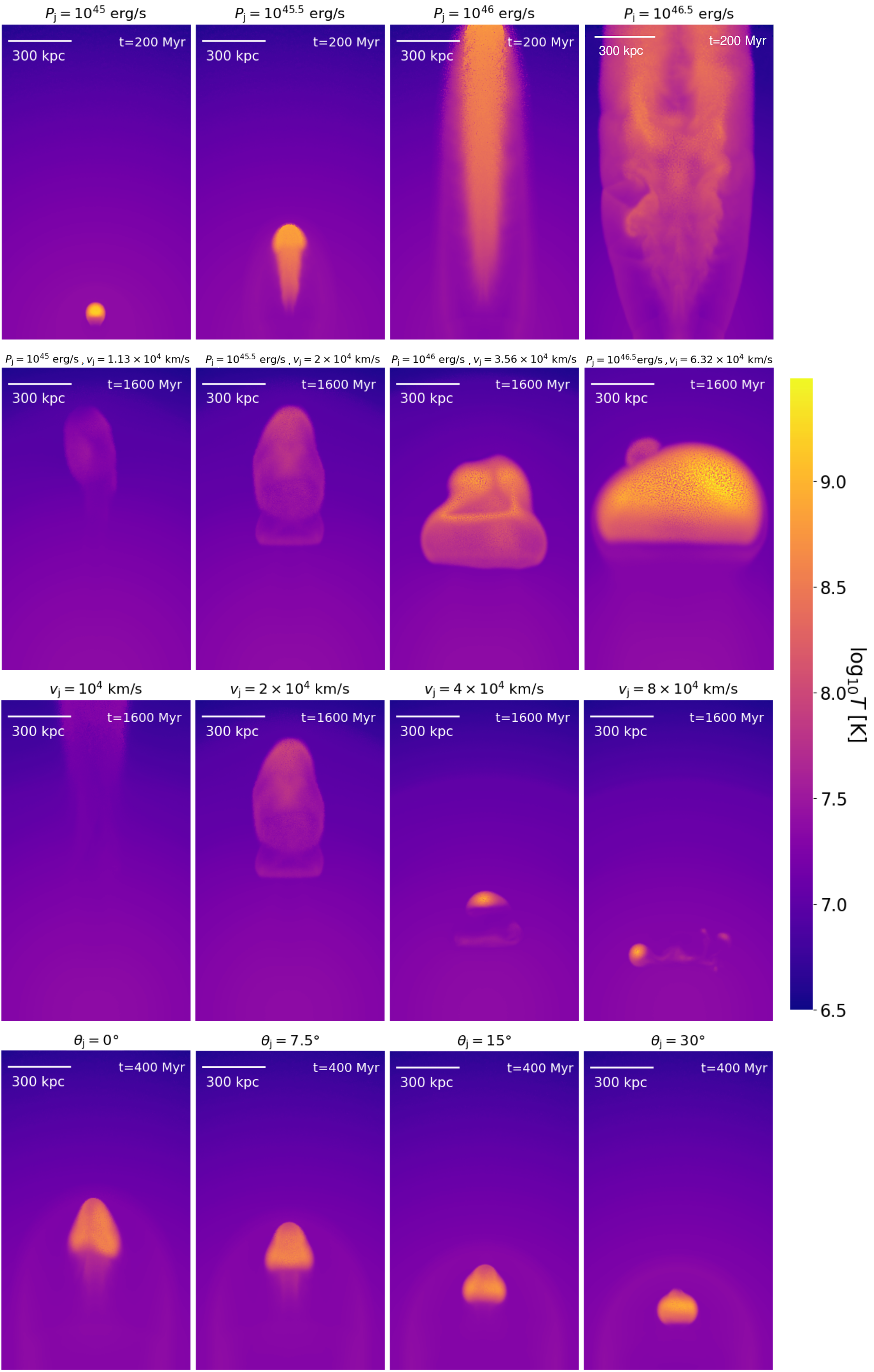}
\caption{Renderings of the gas temperature in simulations with varying physical parameters (first row - jet power, second row - jet power and launching velocity, third row - launching velocity, fourth row - half-opening angle). The standard set of parameters is $P_\mr{j}=3.16\times10^{45}$ ergs$^{-1}$, $v_\mr{j}=2\times10^4$ kms$^{-1}$ and $\theta_\mr{j}=15\degree$. In each case of a varying parameter(s), we show the visualisation at a specific time of interest for that variation (top right corner in each panel). The relevant parameter(s) being varied in each simulation is shown above each panel. The jet duration is 50 Myr, and the numerical resolution $m_\mr{gas}=10^5$ $\mr{M}_\odot$.}
\label{fig:fig9}
\end{figure*}

\begin{figure*}
\includegraphics[width=0.91\textwidth, trim = 0 14 0 0]{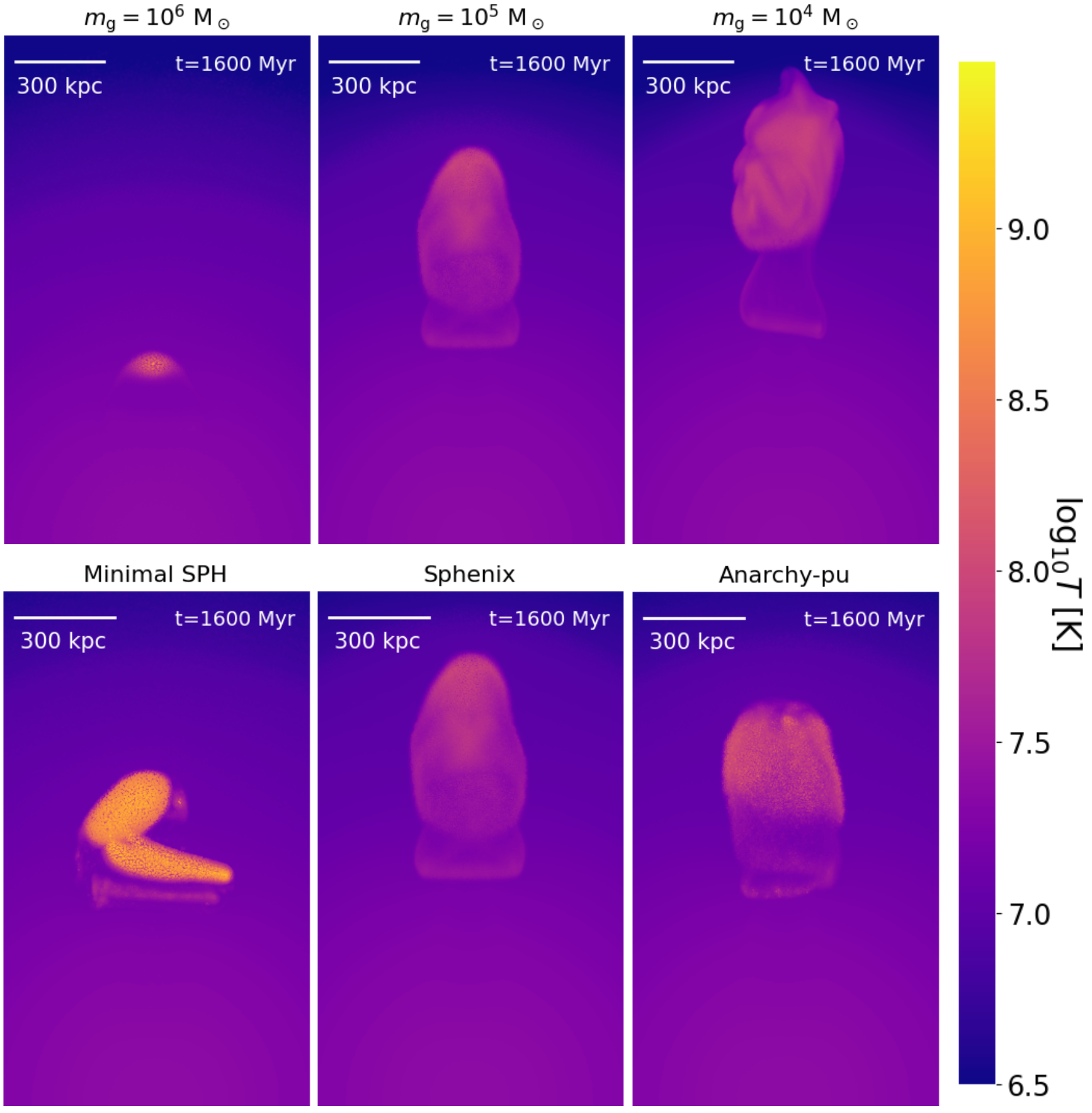}
\caption{Renderings of the gas temperature in simulations with varying numerical resolution (top row) and hydrodynamical scheme (bottom row). The jet-related parameters used in these simulations are $P_\mr{j}=3.16\times10^{45}$ ergs$^{-1}$, $v_\mr{j}=2\times10^4$ kms$^{-1}$ and $\theta_\mr{j}=15\degree$, $T_\mr{j}=50$ Myr. The numerical resolution or scheme being used in each simulation is shown above each panel.}
\label{fig:fig10}
\end{figure*}

We analyse these different simulations through visual renders of the bubbles in terms of temperature, which are shown in Fig. \ref{fig:fig9} for variations of jet-related parameters (jet power, velocity and half-opening angle), as well as in Fig. \ref{fig:fig10} for variations of numerical resolution and hydrodynamical scheme. We have also
studied these different simulations in terms of the energetics, but we find that they are overall similar, and that any differences among the different simulations agree with qualitative differences in the visualisations.

\subsection{Varying the jet power}

We vary jet powers by factors of $\sqrt{10}\approx3.16$, with one lower-power and two higher-power simulations relative to the fiducial one (which has $P_\mr{j}=3.16\times10^{45}$ ergs$^{-1}$). In the first row of panels in Fig. \ref{fig:fig9} we show visualisations of the jets/bubbles in these simulations after $200$ Myr of evolution. The impact of this variation is relatively drastic (compared to the differences for the self-similar regime, where changing the jet power changes the jet lengths weakly: $L_\mr{j}\propto P_\mr{j}^{0.2-0.33}$, see e.g. \citealt{Kaiser2007} or \citealt{Husko2022a}). The bubble reaches a smaller distance in the lowest-power case than in the fiducial case (shown in the second panel), and it does not escape the halo. It is also more spherical, because the jet that created it was shorter and thus traveled for a shorter time in the portion of the gaseous halo with a declining density ($\rho\propto r^{-1.5}$), compared to the constant-density core. The case with $P_\mr{j}=10^{46}$ ergs$^{-1}$ (in the third panel), on the other hand, features a jet that has traveled to $r\approx1500$ kpc by $t=200$ Myr, which is already well outside the virial radius. 

These drastic differences in the three lower-power cases are possibly related to the fact that the mass resolution and launching velocity are the same, so lower jet powers (and total energies) lead to a smaller number of gas particles being launched, and therefore a worse-resolved jet. More likely these differences arise due to lower total jet/bubble masses and momenta, which means that they are more quickly decelerated by the action of the various processes at play (drag, drift, wake, entrainment). In the last snapshot in the first row of Fig. \ref{fig:fig9} we show our highest-power simulation. This jet shows signs of instabilities -- this is a result of increased resolution due to more particles being launched into the jet. The outflow also appears to take a conical and pronged shape. This indicates that the jet was ballistic for most of its evolution, and it is a result of using low velocities in combination with a high power (see \citealt{Husko2022a}).

\subsection{Varying both the jet power and launching velocity}

We now vary the jet powers as in the previous case, but we also vary launching velocities by factors of $10^{1/4}\approx\sqrt{3.16}\approx1.78$, relative to the fiducial choice of $v_\mr{j}=2\times10^4$ kms$^{-1}$. This ensures that the mass injected into the jets remains the same regardless of jet power, and in turn also the numerical resolution (number of particles) in the jets/bubbles. From the second row in Fig. \ref{fig:fig9}, showing the bubbles at $t=1600$ Myr, we see that the differences in bubbles are less drastic than if only the jet power was varied. Somewhat surprisingly, the bubbles reach smaller and smaller distances as the jet power and launching velocity are increased. This is likely due to a backflow that widens the jet lobes at higher launching velocities (\citealt{English2016}, \citealt{Li2018}, \citealt{Husko2022a}), which is visible in the form of wider bubbles shown in the panels. The bubbles are also more spherical with higher velocities$-$this is possibly a result of higher energies per unit mass (initially kinetic as the gas is launched into the jet, but then thermal as it is shocked and becomes part of the lobes). The higher temperatures of the bubble result in a shorter sound-crossing time, therefore allowing for the bubble to expand more uniformly as it rises.


\subsection{Varying the jet launching velocity}

We now discuss cases with varying jet launching velocities, by factors of two relative to our fiducial value of $v_\mr{j}=2\times10^4$ kms$^{-1}$. The visualisations of these simulations are shown in the third row of Fig. \ref{fig:fig9}. As is visible, the jet launching velocity can impact the bubbles significantly. Lower launching velocities lead to more elongated, less spherical bubbles that traverse larger distances. This is likely due to a larger momentum in the jets with lower velocities (the total jet momentum is $p_\mr{j}=2E_\mr{j}/v_\mr{j}$). With the highest launching velocity, the bubble has broken up at a small distance. This is potentially due to lower resolution of the bubble (the number of particles in the bubble scales as $N\propto 1/v_\mr{j}^2$), the backflow at higher launching velocities, or different stability properties of the bubble at smaller distances. These relatively significant differences found by using various values for the jet velocity may also result in different findings for the relevant feedback-related properties such as energetics or gas uplift (discussed at length in Sections \ref{sec:gen_char} and \ref{sec:drift_wake}). However, we note that this is expected, as we consider the jet velocity to be the main parameter in simulations with our numerical model of jet feedback (e.g. \citealt{Husko2022a}, \citealt{Husko2022b}).

\subsection{Varying the jet half-opening angle}

We now discuss cases with varying half-opening angles. Our fiducial choice is $\theta_\mr{j}=15\degree$, and we compare with simulations where it is a factor of two lower and higher, as well as with a case with $\theta_\mr{j}=0\degree$. The visualisations of these simulations are shown in the bottom panels of Fig. \ref{fig:fig9}, at $t=400$ Myr. The bubbles are wider and reach smaller distances with larger opening angles. However, the differences are not drastic. This is likely due to the change in behaviour once the jets leave the core of the gaseous halo. At that point, they begin to expand more freely near the jet head, and the opening angle becomes less significant. 

\subsection{Impact of numerical resolution}

It is important to verify that the simulations converge to the same results if numerical resolution is varied. We compare our standard resolution of $m_\mr{gas}=10^5$ $\mr{M}_\odot$ with simulations that are a factor of 10 higher (our highest-resolution simulation, the one that we discussed in Sections \ref{sec:gen_char} and \ref{sec:drift_wake}) and a factor of 10 lower in mass. The visualisations of the different simulations are shown in the top panels of Fig. \ref{fig:fig10}. The low-resolution bubble appears shorter and has traveled a smaller distance than the two higher-resolution ones, which are very similar in shapes and positions. The low-resolution jet is likely too spherical due to spherical averaging in SPH. The agreement at higher resolutions is encouraging.

\subsection{Impact of hydrodynamical scheme}

The hydrodynamical scheme we used for all of the simulations presented so far was SPHENIX (\citealt{Borrow2022}). It includes an artificial viscosity limiter, since artificial viscosity is a known problem for SPH. It also has artificial conductivity, which is meant to reproduce diffusion of energy through unresolved mixing. Here we compare SPHENIX to a 'minimal SPH' scheme in SWIFT, which mimics traditional SPH schemes without artificial viscosity limiters or artificial conduction (\citealt{Monaghan}). We also compare to anarchy-pu, the scheme used for the EAGLE simulations (\citealt{Schaye2015}, \citealt{Schaller2015}).

The bottom panels of Fig. \ref{fig:fig10} show the bubbles simulated with each of these schemes. The bubble in minimal SPH is more spherical and hotter than the ones in SPHENIX and anarchy-pu, as well as being in the process of breaking up. The bubbles are fairly similar in the other two schemes. The minimal SPH bubble also shows 'droplets' in its wake, which is likely parts of the bubble detaching due to artificial surface tension (\citealt{Argetz2007}, \citealt{Sijacki2012}, \citealt{Nelson2013}), likely from a lack of an artificial viscosity limiter. It is hotter probably due to a lack artificial conduction. The anarchy-pu bubble is not hotter than the SPHENIX one, indicating similar conduction. However, it does feature a somewhat more grainy and less smooth surface, possibly due to a somewhat smaller conduction or viscosity limiter. Overall we find that using more modern SPH schemes leads to what are likely more realistic bubbles.

\section{Conclusions}
\label{sec:conclusions}

We have used the SWIFT smoothed particle hydrodynamics code to simulate bubbles that form in the aftermath of jet episodes launched by active galactic nuclei. The jets are launched with a constant power into a hot, gaseous halo in hydrostatic equilibrium, representing the intracluster medium of a low-mass galaxy cluster ($M_{200}=10^{14}$ $\mathrm{M}_\odot$). We focused on a high-power ($P_\mathrm{j}=3.16\times10^{45}$ $\mathrm{erg}\mathrm{s}^{-1}$), explosive jet episode, rather than gentle, 'maintenance-mode' feedback. In order to resolve the jets and lobes to a sufficient degree, a subrelativistic jet velocity was used ($v_\mathrm{j}=2\times10^4$ $\mathrm{km}\mathrm{s}^{-1}$). With these fiducial parameters, we studied the evolution of the jet-inflated bubbles and their interaction with the ambient intracluster medium as they rise through it. We also performed simulations with varying jet powers, launching velocities, half-opening angles, numerical resolution and hydrodynamical scheme. From our simulations we find the following:

\begin{itemize}
    \item After the jets are turned off, buoyancy begins to act on the shocked jet gas, first from the sides. The shocked jet gas transitions from a jet-like shape to a roughly spherical bubble, eventually developing an indentation in its centre.
    \item As the bubbles rise, they draw out filaments of low-entropy gas from the central regions of the gaseous halo. The filaments are mostly made up of the drift, which is the ICM that was displaced as the bubble passed, and then pulled along behind it. We find that the volume of the drift is roughly 10 per cent the volume of the bubbles, much lower than the value appropriate for a spherical bubble moving through a constant-density medium in the absence of gravity (50 per cent). Despite this, the drift is more massive than the bubbles by roughly an order of magnitude. 
    \item The bubbles are also accompanied by the wake, the ICM trapped in an indentation at the bottom of the bubbles, which travels along with the bubble. The mass in the wake is $15$ per cent of the initially displaced mass (while 24 per cent is expected for spherical bubbles in a constant gravitational field). 
    \item Entrainment of the ICM is important only during the early phase (when the bubbles are forming), where the amount of it is consistent with being caused first by the momentum-driven launching of the jets, and then Rayleigh-Taylor instabilities.
    \item Most of the injected jet energy is transferred to the ICM through bow shocks, with $70\%$ of the energy imparted to the ICM in thermal form, and $30\%$ in kinetic, while the jets are active. Later on, as the wake and drift begin to rise, almost all of the injected energy is in the change of the gravitational potential energy (for both the bubbles and the ICM). This is consistent with changes seen in the radial profiles of gas-related quantities such as density, which indicate that the drawing out of the drift and wake has reduced the central density of the gaseous halo by a factor of two, while the central temperature is raised by 40 per cent. It is possible that, in a realistic scenario, jet feedback proceeds not only through heating of the ICM, but also through a reduction of the central density.
    \item We have compared the evolution of our simulated bubble properties, such as density, radius, velocity and height, with predictions from an analytical model by \cite{Pope}. We find that the prediction of the model with numerical coefficients chosen to match the ones we measured (discussed above) reproduces the simulated behaviour if the drag coefficient is $C_\mathrm{D}=0.35$, close to the value appropriate for a bullet-shaped object at large Reynolds numbers ($C_\mathrm{D}=0.3$). Alternatively, the simulated bubbles are also consistent with drag dominating over the drift, wake and entrainment, all of which are negligible. In this picture, however, the drag coefficient is time-dependent and evolves from small values ($C_\mathrm{D}<0.15$) to values appropriate for a spherical bubble at late times ($C_\mathrm{D}=0.55$).
    \item By varying the jet power and jet velocity, as well as both at the same time, we find that the dominant changes in late-time behaviour of the bubbles do not come from extra energy, but rather extra momentum of the jets. Bubbles with more momentum can travel much farther in the ICM, to the point of escaping it. The jet velocity strongly impacts the shape and stability of the bubbles, as well as their time evolution (in the sense of distance traveled or velocity vs. time). By varying the half-opening angle of the jets, we find more spherical bubbles that travel shorter distances for larger opening angles, as expected. By varying numerical resolution, we find that our simulations are well converged. At our standard resolution, comparing bubbles simulated using the SPHENIX SPH scheme with those simulated using minimal SPH, we find that minimal SPH has too much artificial viscosity and not enough diffusion. The anarchy-pu scheme, which was used for the EAGLE simulations, matches our bubbles closely.
\end{itemize}

We again stress that the above conclusions do not necessarily hold for all AGN jet feedback events, since we simulated a fairly powerful episode. From a more qualitative side, however, we expect them to remain true.  In future work we will study jets and the bubbles they inflate in more realistic environments (e.g. zoom-in cosmological simulations of galaxy clusters). We will perform simulations of self-consistent and spin-driven jet feedback, so that the jet power varies with time in a realistic manner. These changes may have a significant effect on the interplay between the bubbles and the intracluster medium.

\section*{Acknowledgements}
F.H. would like to thank Matthieu Schaller and Joop Schaye for useful discussions and comments. The research in this paper made use of the SWIFT open-source simulation code (\url{http://www.swiftsim.com}, \citealt{Schaller2018})
version 0.9.0. The swiftsimio Python library was used to analyze and visualize the data from the simulations (\citealt{Borrow2020_swiftsimio}, \citealt{Borrow_2021_swiftsimio}). F. H. would like to acknowledge support from the Science Technology Facilities Council through a CDT studentship (ST/P006744/1), and the STFC consolidated grant ST/T000244/1. This work used the DiRAC@Durham facility managed by the Institute for Computational Cosmology on behalf of the STFC DiRAC HPC Facility (www.dirac.ac.uk). The equipment was funded by BEIS capital funding via STFC capital grants ST/K00042X/1, ST/P002293/1, ST/R002371/1 and ST/S002502/1, Durham University and STFC operations grant ST/R000832/1. DiRAC is part of the National e-Infrastructure.

\section*{Data availability}

The data underlying this article will be provided upon reasonable request to the corresponding author.




\bibliographystyle{mnras}
\bibliography{jet_bibliography} 

\bsp	
\label{lastpage}
\end{document}